\documentclass[reqno,10pt,a4paper]{article}
\usepackage{amssymb}
\usepackage{amsmath}
\usepackage{amsthm}
\usepackage{graphicx}
\usepackage[latin1]{inputenc}
\usepackage{epsf,pstricks}

\DeclareMathAlphabet\RsfsCal{U}{rsfs}{m}{n}
\SetMathAlphabet\RsfsCal{bold}{U}{rsfs}{b}{n}

\newcommand{\mt}[1]{\text{\rm #1}} 
\DeclareMathOperator{\sign}{\mt{sign}}

\numberwithin{equation}{section}
\newtheorem{theorem}{Theorem}[section]
     \newtheorem{lemma}[theorem]{Lemma}
     
     \newtheorem{corollary}[theorem]{Corollary}
     \newenvironment{Proof.}[1][Proof.]{\begin{trivlist}
     \item[\hskip \labelsep {\bfseries #1}]}{\end{trivlist}}

     \newenvironment{acknowledgment}[1][Acknowledgments:]{\begin{trivlist}
     \item[\hskip \labelsep {\bfseries #1}]}{\end{trivlist}}
\theoremstyle{remark}
\newtheorem{remark}[theorem]{Remark}
\newtheorem*{thm1}{Theorem 1}

\newcommand{\field}[1]{\mathbb{#1}}
\newcommand{\C}{\field{C}}
\newcommand{\R}{\field{R}}
\newcommand{\N}{\field{N}}
\newcommand{\Z}{\field{Z}}

\renewcommand{\Im}{\,\mathrm{Im}\,}
\renewcommand{\Re}{\,\mathrm{Re}\,}

\allowdisplaybreaks

\title{Decay Rates for Spherical Scalar Waves in the
Schwarzschild Geometry}
\author{Johann Kronthaler\thanks{Research supported in part by the Deutsche
Forschungsgemeinschaft.}}
\date{September 2007}

\begin{document}
\maketitle

\begin{abstract}
The Cauchy problem is considered for the scalar wave equation in the
Schwarzschild geometry. Using an integral spectral representation
we derive the exact decay rate for solutions of the Cauchy problem with spherical symmetric initial data, which is smooth and compactly supported outside the event horizon.
\end{abstract}


\section{Introduction}
In this paper we study the decay rate for solutions of the scalar wave equation with spherical symmetric initial data in the Schwarzschild geometry, which is smooth and compactly supported outside the event horizon. We prove that these solutions decay at the rate $t^{-3}$ and $t^{-4}$ for momentarily stationary initial data, respectively, as it was earlier predicted by Price \cite{Price}, though not rigorously proved. In \cite{Kronthaler} we have already shown pointwise decay for solutions of the same kind of initial data not necessarily spherical symmetric. To this end, we have derived an integral spectral representation for the solutions applying Hilbert space methods in terms of special solutions of the Schr\"odinger equation the so-called Jost solutions.

In order to set up some notation, recall that in Schwarzschild coordinates $(t,r,\vartheta,\varphi)$,
the Schwarzschild metric takes the form
\begin{eqnarray} \label{schwarzschildgeometrie}
ds^2 & \hspace{-3mm} = \hspace{-2mm} & g_{ij} \: dx^{i} dx^{j} \nonumber \\
 & \hspace{-3mm} = \hspace{-2mm} & \left(1- \frac{2M}{r} \right)\: dt^2 -
\left(1-
 \frac{2M}{r}\right)^{-1} dr^2 - r^2(d\vartheta ^2
+ \mathrm{sin}^2 \vartheta \: d\varphi^2)
\end{eqnarray}
with $r>0,\: 0 \leq \vartheta \leq \pi ,\: 0 \leq \varphi < 2\pi$. The metric has two singularities at $r=0$ and $r=2M$. The latter is called the \textit{event horizon} and can be resolved by a coordinate
transformation. We consider the scalar wave equation in the region $ r> 2M $
outside the event horizon, which is given by
\begin{equation} \label{eq: Wellengleichung Grundform}
\square \phi := g^{ij} \nabla_{i} \nabla_{j} \phi =
\frac{1}{\sqrt{-g}} \frac{\partial}{\partial x^{i}} \left( \sqrt{-g} \:g^{ij}
\frac{\partial }{\partial x^{j}} \right) \phi = 0
\end{equation} where $g$ denotes the determinant of the metric $g_{ij}$.
We now state our main result.
\begin{thm1}
Consider the Cauchy problem of the scalar wave equation in the Schwarzschild
geometry
$$ \square \phi = 0\; , \quad (\phi_0, i \partial_t \phi_0)(0,r,x) = \Phi_0(r,x)$$
for smooth spherical symmetric initial data $\Phi_0 \in C^{\infty}_0 ( (2M,\infty)
\times S^2)^2 $ which is compactly supported outside the event horizon. 
Let $\Phi (t) = (\phi (t), i \partial_t \phi (t)) \in C^\infty(\R \times (2M, \infty)
\times S^2)^2$ be the unique global solution which is compactly supported for all
times $t$. Then for fixed $r$ there is a constant $c= c(r,\Phi_0)$
such that for large $t$
\begin{equation*}
|\phi(t)| \leq \frac{c}{t^3} \; .
\end{equation*}
Moreover, if we have initially momentarily static initial data, i.e. $\partial_t
\phi_0 \equiv 0$, the solution $\phi(t)$ satisfies
\begin{equation*} 
|\phi(t)| \leq \frac{c}{t^4} \; .
\end{equation*}
\end{thm1}

There has been significant work in the study of linear hyperbolic equations in black hole spacetimes.  The first major
contribution in this topic was made in
1957, when Regge and Wheeler studied the linearized equations for
perturbations of the Schwarzschild metric \cite{ReggeWheeler}. This work was
continued in \cite{Visch,Zerilli}, while more
recently the decay of the perturbation and all of its derivatives
was shown in \cite{Friedman} using a theorem by Wilcox. By heuristic
arguments, in 1972 Price \cite{Price} got evidence for polynomial decay of solutions
of the scalar wave equation in Schwarzschild, where the power depends explicitely on
the
angular mode.
In 1973, Teukolsky \cite{Teu1} could derive by means of the Newman Penrose formalism
one single master equation that describes in the Kerr background the evolution of a
test scalar field ($s=0$), a test neutrino field ($s=\pm1/2$), a
test electromagnetic field ($s= \pm 1$) and linearized gravitational waves
($s= \pm 2$).
Here, the parameter $s$ is also called the spin weight of the field. Note that
it is a quite complicated task in the case $s \neq 0$ to recover all the components
of the corresponding field from a solution of
this equation. For further details see
\cite{Cha,Whiting1}. In two subsequent papers
\cite{Teu2,Teu3}, Teukolsky and Press discussed the physical consequences of
these perturbations. Note that the rigorous analysis of the equation remains a quite subtle point, though any linearized perturbation is given by this equation.
For instance, in the case $s\neq0$ \textit{complex} coefficients are involved, which
makes the analysis very complicated. Hence, until now there are just a few
rigorous results in this case. In \cite{Finster3} local decay was proven for the
Dirac equation ($s=\frac{1}{2}$) in the Kerr geometry (in the massless and massive
case). Moreover, a
precise decay rate has been specified in the massive case \cite{Finster4}. More
recently, there has been a linear stability result for the Schwarzschild geometry
under electromagnetic and gravitational perturbations \cite{Finster5}. This result
relies on the mode analysis, which has been carried out in \cite{Whiting2}. More
work has been done on the case $s=0$, where the Teukolsky equation reduces to
the scalar wave equation. In the Schwarzschild case, Kay and Wald \cite{KayWald}
proved a time independent $L^\infty$-bound for solutions of the Klein-Gordon
equation. In \cite{Dafermos}, a mathematical proof is given for the decay rate of
solutions with spherical symmetric initial data, which has been predicted by Price
\cite{Price}, which is not sharp, however. For general initial data, the same authors derived another decay result
\cite{Dafermos2}. Pointwise decay in the Kerr geometry was proven rigorously \cite{Finster1,Finster2}. Furthermore, Morawetz and
Strichartz-type estimates for a massless scalar field without charge in a
Reissner Nordstr{\o}m background with naked singularity are developed in
\cite{Stalker}. And in \cite{Blue} a Morawetz-type inequality was proven for the
semi-linear wave equation in Schwarzschild, which is also supposed to yield decay
rates.

In this paper we first recapitulate the framework and some notations of the foregoing paper \cite{Kronthaler}. Afterwards, we give an explicit expansion of the Jost solutions $\grave{\phi}$ of the Schr\"odinger equation which also were derived in \cite{Kronthaler}. At the end we show how to derive the exact decay rate out of this expansion.

\section{Preliminaries}
We usually replace the Schwarzschild radius $r$ by the Regge-Wheeler coordinate $u \in \R$ given by
\begin{equation} \label{reggewheelercoord}
u(r) := r + 2M\: \mathrm{log} \left(\frac{r}{2M} -1 \right) \: .
\end{equation}
After having separated the angular modes $l,m$ using spherical harmonics, it is convenient to write the Cauchy problem in Hamiltonian formalism
\begin{equation} \label{Hamiltonform allgemein}
i \partial_t \Psi = H \Psi \; , \quad \Psi \big|_{t=0} = \Psi_0
\end{equation}
where $\Psi=(\psi,i\partial_t \psi)^T$ is a two component vector representing the wave function and its first time derivative and $H$ is the Hamiltonian
\begin{equation} \label{Hamiltonian allgemein}
\left(%
\begin{array}{cr}
  0 & 1 \\
 -\partial_u^2 + V_l(u) & 0 \\
\end{array}%
\right) \; ,
\end{equation}
with the potential
\begin{equation} \label{potential}
V_l(u) = \left( 1- \frac{2M}{r} \right) \left(\frac{2M}{r^3} +
\frac{l(l+1)}{r^2}
\right) \: .
\end{equation}
Constructing the resolvent of the operator $H$ and using Stone's formula we have derived an integral spectral representation for the solutions of the Cauchy problem of the following form
\begin{eqnarray}
 \Psi(t,u)=e^{-itH}
\Psi_0 (u) =\hspace*{70mm} \nonumber \\ - \frac{1}{\pi}   \int_{\mathbb{R}} e^{-i \omega t}
\left(
\int_{\textrm{supp}\: \Psi_0} \mathrm{Im} \! \left(
\frac{\acute{\phi}_{\omega l}(u)
\grave{\phi}_{\omega l}(v)}{w(\acute{\phi}_{\omega l},\grave{\phi}_{\omega
l})}\right)
 \left(%
\begin{array}{lc}
  \omega & 1 \\
  \omega^2 & \omega \\
\end{array}%
\right)  \Psi_0(v) dv \right)  \: d\omega \; , \label{eq: Darstellung der Loesung l=0}
\end{eqnarray}
where the integrand is in $L^1$ with respect to $\omega$. 

At this point, the functions $\acute{\phi},\grave{\phi}$ play an important role. These functions are a fundamental system of the Schr\"odinger equation
\begin{equation} \label{Schroedinger equation}
\left( -\partial_u^2  +V_\omega(u) \right) \phi(u) = 0
\end{equation}
with the potential
\begin{equation} \label{schroedinger potential}
V_\omega(u) = - \omega^2 + V_l(u) = - \omega^2 + \left( 1- \frac{2M}{r} \right)
\left(\frac{2M}{r^3} + \frac{l(l+1)}{r^2} \right)
\end{equation}
with boundary conditions
\begin{eqnarray}
\lim_{u \rightarrow -\infty}  e^{-i \omega u} \acute{\phi}_{\omega}(u) =1 \; ,
\quad & \displaystyle \lim_{u \rightarrow -\infty}  \left( e^{-i \omega u}
\acute{\phi}_{\omega}(u) \right)' =
0  \label{boundarycond3} \\
\lim_{u \rightarrow +\infty}  e^{i \omega u} \grave{\phi}_{\omega}(u) =1 \; ,
\quad & \displaystyle \hspace{4mm} \lim_{u \rightarrow +\infty}  \left( e^{i
\omega u}
\grave{\phi}_{\omega}(u) \right)' = 0 \; . \label{boundarycond4}
\end{eqnarray}
in the case $\Im (\omega) <0$. We derived these solutions using the corresponding integral equation, the so-called Jost equation, which is given by 
\begin{equation} \label{Jost equation bound cond -unendlich}
\phi_\omega (u) = e^{i \omega u} + \int _{-\infty}^u \frac{1}{\omega} \sin(\omega(u-v)) V_l(v)
\phi_\omega(v) \:dv \; ,
\end{equation}
in the case of boundary conditions at $-\infty$ (an analog equation is considered for the boundary conditions at $\infty$). Now, we have constructed the solution $\acute{\phi}$ with the series ansatz 
\begin{equation} \label{Reihenentwicklung von phi 1}
\acute{\phi}_\omega = \sum_{k=0}^\infty \phi_\omega^{(k)} \; ,
\end{equation}
together with the iteration scheme
\begin{equation} \label{Iterationsschema fuer phi 1}
\left. \begin{array}{rcl}
  \phi_\omega^{(0)} (u) & = & e^{i \omega u} \\
  \ & \vdots & \ \\
    \phi_\omega^{(k+1)}(u) & = & \displaystyle  \int_{-\infty}^u \frac{1}{\omega} \sin(\omega(u-v))
V_l(v) \phi_\omega^{(k)}(v) \:dv \\
\end{array} \right\}  \begin{array}{c}
   \\
   \\
  . \\
\end{array}
\end{equation}
Using this we have proven that the solutions $\acute{\phi}_\omega(u)$ are analytic with respect to $\omega$ for fixed $u$ in the region $\Im(\omega)<0$. Moreover $\acute{\phi}_\omega$ can be analytically extended to the region $\Im (\omega) \leq \frac{1}{4M}$, whereas the solution $\omega^l \grave{\phi}_\omega$ can be only continuously extended to the real axis. Thus, in order to obtain the exact decay rates it is important to analyze the behavior of $\grave{\phi}_\omega$ with respect to $\omega$ at the real axis in more details.

\section{Expansion of the Jost solutions $\grave{\phi}_\omega$}

Since the $\omega$-dependence of the Jost solutions $\grave{\phi}_\omega$ plays an
essential role in the analysis of the integral representation, we show in this
section a method to expand these solutions at the critical point $\omega
=0$. We start with an explicit calculation:

\begin{lemma} \label{lemma: zur Reihenentwicklung}
For all $u>0$, $\omega \in \R \setminus \{0\}$, $\varepsilon >0$, $q \in \N_0$ and $p
\in \N$,
\begin{eqnarray}
\int_u^\infty e^{-2 i \omega x - \varepsilon x}  \frac{\log^q(x)}{x^p}\: dx = 
\sum \limits_{m=0}^q \begin{pmatrix}
 q\\ m
\end{pmatrix} \log^{q-m} (u) \nonumber \Bigg\{ \left(2 i \omega+\varepsilon
\right)^{p-1} \hspace*{1cm} \\ \hspace*{2mm} \times \left[
\frac{(-1)^{p-1}}{(p-1)!} \frac{(-1)^{m+1}}{m+1} \log^{m+1}\left[(2 i
\omega+\varepsilon) u\right] + \sum \limits_{k=0}^m c_k(m) \log^k \left[(2 i
\omega+\varepsilon) u\right]\right] \nonumber\\
 - u^{-p+1}\sum \limits_{k=0,k\neq p-1}^\infty \frac{(-1)^k (-1)^m \:
m!}{(k-p+1)^{m+1}
k!} \left[(2 i \omega+\varepsilon) u\right]^k  \Bigg\} \;, \hspace*{10mm} \label{eq:
Lemma zur
Reihenentwicklung}
\end{eqnarray}
where the coefficients $c_k$ involve the coefficients $a_0,...,a_q$ of the series
expansion of the $\Gamma$-function at $1-p$.
\end{lemma}

\begin{proof}
In order to prove this, we write the integral as $\lambda$-derivatives,
\begin{equation} \label{eq: Umschreiben der log mit Funktional} \int_u^\infty e^{-2 i
\omega x - \varepsilon x} \frac{\log^q(x)}{x^p}\: dx =
\frac{d^q}{d\lambda^q} F_p(\lambda)\bigg|_{\lambda = 0}\; ,
\end{equation}
with the generating functional,
$$F_p(\lambda) = \int _u^\infty e^{(-2 i \omega -\varepsilon) x } \frac{1}{x^{p-
\lambda}} \: dx = u^{-p+\lambda+1} \int_1^\infty e^{(-2 i \omega- \varepsilon)u
v} \frac{1}{v^{p - \lambda}} \: dv \; ,$$
where in the last step we introduced the new integration variable $v = \frac{x}{u}$.
In the following we will write $z = (2 i \omega
+\varepsilon)u$ for reasons of convenience.
The integral on the right hand side is also known as the Exponential Integral
$E_{p-\lambda} (z)$ with the series
expansion
$$ E_{p- \lambda} (z) = \Gamma( 1- p +\lambda) \: z^{p- 
\lambda -1}- \sum \limits _{k=0}^\infty \frac{(-1)^k}{(k-p +\lambda +1)k!}\:
z^k \; ,$$ for small $\lambda \neq 0$ [as a reference cf. \cite{Wo}]. Using the
series expansion of the $\Gamma$-function at $1-p \in \Z \setminus \N$, where the
$\Gamma$-function has a
pole
of first-order, we obtain
\begin{eqnarray}
 F_p(\lambda) = u^{-p+\lambda+1} \left[ \left( \frac{(-1)^{p-1}}{(p-1)! \: \lambda} 
 + \sum \limits
_{n=0}^\infty a_n \lambda^n \right) z^{p-\lambda -1} \right. \hspace*{3cm}
\nonumber\\
 \left. - \sum \limits _{k=0}^\infty
\frac{(-1)^k}{(k-p +\lambda +1)k!}\: z^k \right] \nonumber\\
= u^{-p+\lambda+1}\left[ z^{p-1} \left( \frac{(-1)^{p-1} }{(p-1)!} \left( \frac{z^{-
\lambda} - 1}{\lambda} \right) + z^{-\lambda} \sum \limits _{n=0}^\infty a_n
\lambda^n \right)\hspace*{10mm} \right. \nonumber \\ 
\hspace*{1cm} \left. - \sum \limits _{k=0 , k \neq p-1}^\infty \frac{(-1)^k}{(k-p
+\lambda +1)k!}\: z^k \right]\;. \label{eq: Reihendarstellung fuer Funktional}
\end{eqnarray}
Using $z^{-\lambda} = e^{-\lambda \log z}$, we immediately get the formulas
\begin{eqnarray*}
 \frac{d^n}{d \lambda^n} \left( \frac{z^{-\lambda}-1}{\lambda} \right)\bigg|_{\lambda
= 0} & = &\frac{(-1)^{n+1}
\log^{n+1}(z)}{n+1} \\ \frac{d^m}{d \lambda^m}  \left( u^\lambda \right)
\bigg|_{\lambda = 0} & =& \log^m (u) \\
\frac{d^m}{d \lambda^m}  \left( z^{-\lambda} \right)
\bigg|_{\lambda = 0} & =& (-1)^m \log^m (z)
\end{eqnarray*}
one directly verifies the claim setting (\ref{eq: Reihendarstellung fuer
Funktional}) in (\ref{eq: Umschreiben der log mit Funktional}).
\end{proof}

Directly in the same way, one proves an analogue lemma for the case $ p \in \Z
\setminus \N$:
\begin{lemma} \label{lemma: zur Reihenentwicklung mit p<0}
For all $u>0$, $\omega \in \R \setminus \{0\}$, $\varepsilon >0$, $q \in \N_0$ and $p
\in \Z \setminus \N$,
\begin{eqnarray}
\int_u^\infty e^{-2 i \omega x - \varepsilon x}  \frac{\log^q(x)}{x^p}\: dx =
\hspace*{70mm}
\nonumber\\ 
\sum \limits_{m=0}^q \begin{pmatrix}
 q\\ m
\end{pmatrix} \log^{q-m} (u) \nonumber \Bigg\{ \left(2 i \omega+\varepsilon
\right)^{p-1}  \sum \limits_{k=0}^m c_k(m) \log^k \left[(2 i
\omega+\varepsilon) u\right] \hspace*{5mm} \\- u^{-p+1}\sum \limits_{k=0}^\infty
\frac{(-1)^k (-1)^m \:
m!}{(k-p+1)^{m+1}
k!} \left[(2 i \omega+\varepsilon) u\right]^k  \Bigg\}  \label{eq: Lemma zur
Reihenentwicklung mit p<0}
\end{eqnarray}
where the coefficients $c_k$ involve the coefficients $a_0,...,a_q$ of the series
expansion of the $\Gamma$-function at $1-p$.
\end{lemma}
Compared to Lemma \ref{lemma: zur Reihenentwicklung}, here the logarithmic term is
of lower order due to the fact that the Gamma-function has no singularity for
positive integers. 

In order to apply this lemma to our integral representation, we have to derive an
asymptotic expansion for the potential $V_l(u)$ at $+ \infty$. Therefore, we have the
following 

\begin{lemma} \label{lemma: asymptotische Entwicklung von V_l}
For the potential $\displaystyle V_l(u) = \left(1- \frac{2M}{r(u)} \right) \left(
\frac{2M}{r(u)^3} + \frac{l(l+1)}{r(u)^2} \right)$ we have the
asymptotic expansion
\begin{equation} \label{eq: asymptotische Entwicklung von V_l}
V_l(u) = \sum \limits _{p=2}^k \sum \limits _{q=0}^{p-2} c_{pq} \frac{\log^q
(u)}{u^p} + c_{k+1,k-1} \frac{\log^{k-1}(u)}{u^{k+1}} +  \mathcal{O}
\left(\frac{\log^{k-2}(u)}{u^{k+1}}\right) \; , 
\end{equation}
as $ u \rightarrow \infty$ ,
with $k \geq 2$ and real coefficients $c_{pq}$, where e.g. the first coefficients
are given by
$$
\begin{array} {l}
c_{20} = l(l+1) \; , \quad c_{31} = 4 l (l+1) M \; , \\ 
\vspace*{-2mm} \\
c_{30} = 2 M  -  2M l (l+1) (1 + 2\log(2)) - 4M l (l+1) \log(M) \\
\vspace*{-2mm} \\
c_{42} = 12 l (l+1) M^2 \; , \\ 
\vspace*{-2mm} \\
c_{41} =-4 M^2(-3 + l(l+1)(5 + 8 \log(8)) + 6 l(l+1)\log(M)) \; ,\;...\;\;.
\end{array}
$$
Furthermore, in the case $l=0$ the coefficients $c_{n,n-2}$ vanish.
\end{lemma}

\begin{proof}
First we have to find an expression for $r$ in terms of the
Regge-Wheeler coordinate $u$. Remember that
$ u = r + 2 M \log (\frac{r}{2M}-1)$, which is equivalent to 
$$ e^{\frac{u}{2M}-1} = \left( \frac{r}{2M}-1 \right)e^{\frac{r}{2M}-1}\; .$$
In order to resolve this equation with respect to $r$, we use the principal branch of
the Lambert $W$ function denoted by $W(z)$. This is just the inverse function of
$f(x)
=x e^x$ on the positive real axis. [As a reference cf. \cite{LambertW}.] Hence, we
obtain
\begin{equation} \label{eq: r ausgedrueckt durch u}
r = 2M +2M \: W(e^{\frac{u}{2M}-1}) \; .
\end{equation}
Moreover, for $W$ we have the asymptotic expansion
\begin{equation} \label{eq: asymptotische Entwicklung der LambertW}
W(z) = \log z -\log(\log z) + \sum \limits_{k=0}^\infty \sum \limits_{m=0}^\infty
c_{km} \: (\log (\log z))^{m+1} (\log z)^{-k-m-1} \; ,
\end{equation}
as $ z \rightarrow \infty$. Here, the coefficients $c_{km}$ are given by $c_{km}
= \frac{1}{m!} (-1)^k \begin{bmatrix}
 k +m  \\ 
 k+1
\end{bmatrix}$, where $ \begin{bmatrix}
 k +m  \\ 
 k+1
\end{bmatrix} $ is a Stirling cycle number. In particular, applying this expansion to
(\ref{eq: r ausgedrueckt durch u}), we get the series representation 
\begin{eqnarray*}
r(u) =2M +2M\Big[ \frac{u}{2M}-1 -\log\left(\frac{u}{2M}-1\right) \hspace*{3cm}  \\
 + \sum \limits_{k=0}^\infty \sum \limits_{m=0}^\infty
c_{km} \: \left(\log \left(\frac{u}{2M}-1\right)\right)^{m+1}
\left(\frac{u}{2M}-1\right)^{-k-m-1}\Big] \; .
\end{eqnarray*}
This allows us to expand the powers $\frac{1}{r^2}, \frac{1}{r^3}$ and
$\frac{1}{r^4}$ to any order in $u/2M -1$
using the method of the geometric series. Together with the expansion 
$$ \log \left(\frac{u}{2M}-1\right) = \log \left[\frac{u}{2M}\left(1
-\frac{2M}{u}\right) \right] = \log u -\log (2M) -\sum \limits _{n=1}^\infty
\frac{1}{n} \left( \frac{2M}{u} \right)^n , $$ which holds for $u>2M$, the result
follows.
\end{proof}

These two lemmas let us expand the solution $\grave{\phi}_\omega(u)$ in the
following way.


\begin{lemma} \label{lemma: Entwicklung von acute phi,l=0}
For $l=0$, $\omega \in \R \setminus \{0\}$ and fixed $u>0$, the fundamental solution
$\grave{\phi}_\omega(u)$ can be represented as
\begin{equation} \label{eq: Entwicklung von acute phi,l=0}
\grave{\phi}_\omega(u)= e^{-i \omega u} + g_0(\omega,u) + 2 i \omega \log(2 i \omega)
g_1(\omega,u) +2 i \omega g_2(\omega,u) \; ,
\end{equation} 
where the functions $g_0,g_1$ and $g_2$ are $C^1(\R)$ with respect to $\omega$.
\end{lemma}

In order to prove this, we need the following lemma

\begin{lemma} \label{lemma: abschaetzung der omegaabl der greensfkt}
For all $u \in \C$ and $ n\in \N_0$,
\begin{equation} \label{eq: absch von ableitung von sinus}
\Big| \partial^n_u \left(\frac{1}{u} \sin u \right) \Big| \leq \frac{2^{n+1}}{1 +
|u|}
e^{|\Im u|} \; .
\end{equation}
Moreover, if $\omega \neq 0$ and $v \geq u >0$,
\begin{equation} \label{eq: abschaetzung der omegaabl der greensfkt}
\Big| \partial^n_\omega  \big[ \frac{1}{\omega} \sin(\omega(u-v))\big] \Big| \leq
\frac{C(n) \: v^{n+1}}{1 +| \omega v|} e^{v |\Im \omega| + u \Im \omega} \; ,
\end{equation}
for some constant $C(n)$, which is just depending on $n$.
\end{lemma}

\begin{proof}
In the case
$|u| \geq
1$, Eulers formula for the 
$\sin$-function  
inductively yields
$$ (1 +|u|) \Big| \partial^n_u \left( \frac{1}{u} \sin u \right) \Big| 
\leq 2^{n+1} e^{| \Im u|} \; .$$

For $|u|<1$ we rewrite $ (1/u) \sin u$ as an integral, in order to obtain the
estimate
$$(1 +|u|)\Big| \partial^n_u \left( \frac{1}{u} \sin u \right) \Big| =  (1 + |u|)
\Big|
\frac{1}{2} \int_{-1}^1 (i \tau)^n e^{i u \tau} \: d \tau \Big |  \leq 2 e^{|\Im
u|}\;
,$$ which shows the first claim. As a consequence, we get for $ \omega \neq 0$ and
all $n \in \N$ the estimate
\begin{eqnarray} \nonumber 
 \Big|  \partial^n_\omega \left( \frac{1}{\omega} \sin (\omega u) \right)
\Big| & = & \Big| u^{n+1} \partial^n_{\omega u} \left(\frac{1}{\omega u} \sin (\omega
u)
\right) \Big| \\ & \leq & \frac{2^{n+1} |u|^{n+1}}{1 + |\omega u|} e^{|\Im (\omega
u)|} \;
. \label{eq: abschaetzung greensfkt zwischenschritt}
\end{eqnarray}

In order to show (\ref{eq: abschaetzung der omegaabl der greensfkt}) we use the identity
\begin{equation} \label{identity} \frac{1}{\omega} \sin(\omega(u-v)) =  \frac{1}{\omega} \left(
\sin (\omega u) e^{ i \omega v} - \sin (\omega v) e^{i \omega u} \right)
\end{equation}
and apply (\ref{eq: absch von ableitung von sinus}), $n=0$,
\begin{eqnarray}
 \bigg| \frac{1}{\omega} \sin(\omega(u-v))
\bigg| \: \leq \: \frac{1}{| \omega |}  \left( \big|\sin (\omega u) e^{ i \omega
v}
\big| +
\big|\sin (\omega v) e^{i \omega u} \big| \right) \nonumber\\
\leq  \frac{2 |u|}{1+
|\omega u|} \,  e^{|u \Im \omega|} e^{-v \Im \omega}  + \frac{2 |v|}{1+ |\omega
v|} \, e^{|v \Im \omega|} e^{-u \Im \omega} \; . \label{eq: Zwischenschritt im
Lemma zur abschaetzung von G}
\end{eqnarray}
Due to the assumption $v\geq u \geq 0$, we know that $|v| \geq |u|$ and thus
$$ \frac{2 |u|}{1+ |\omega u|} \leq \frac{2 |v|}{1+ |\omega v|}\; , \quad
 \;  u | \Im \omega | + v \Im \omega \geq   v | \Im \omega | + u \Im \omega \;
. $$
Using these inequalities in (\ref{eq: Zwischenschritt im
Lemma zur abschaetzung von G}) the claim follows for $n=0$.

Once again using the identity \ref{identity}
we get
\begin{eqnarray*}
\Big| \partial_\omega \left( \frac{1}{\omega} \sin(\omega(u-v)) \right)
\Big| \leq
\Big| \frac{1}{\omega} \left( \sin (\omega u) (-iv)  e^{ - i \omega v} - \sin
(\omega v) (-i u) e^{-i \omega u} \right) \Big|  \\ 
+ \:\Big| \partial_\omega \left(\frac{1}{\omega} \sin (\omega u) \right) e^{ - i
\omega v} -  \partial_\omega \left(  \frac{1}{\omega} \sin (\omega v) \right) e^{-i
\omega u} \Big|\; .
\end{eqnarray*}
Using the estimates (\ref{eq: abschaetzung greensfkt zwischenschritt}) and (\ref{eq:
absch von ableitung von sinus}) for $n=0$ together with the assumption $v \geq u>0$,
we see as before that the
first term is bounded by
$$ \Big| \frac{1}{\omega} \left( \sin (\omega u) (-iv)  e^{ - i \omega v} - \sin
(\omega v) (-i u) e^{-i \omega u} \right) \Big| \leq \frac{4 v^2}{1+ | \omega v|} e^{
v |\Im \omega| + u \Im \omega} \;. $$ 
For the second term we use (\ref{eq: abschaetzung greensfkt zwischenschritt})
\begin{eqnarray*}
\Big| \partial_\omega \left(\frac{1}{\omega} \sin (\omega u) \right) e^{ - i
\omega v} -  \partial_\omega \left(  \frac{1}{\omega} \sin (\omega v) \right) e^{-i
\omega u} \Big| \hspace*{20mm}\\ 
\leq \frac{4u^2}{1 + |\omega u|} e^{u | \Im \omega | + v \Im
\omega} +\frac{4v^2}{1 + |\omega v|} e^{v | \Im \omega| + u \Im \omega} \; ,
\end{eqnarray*} and obtain due to the assumption $v\geq u>0$
$$\leq \frac{8 v^2}{1+ | \omega v|} e^{ v |\Im \omega|+u \Im \omega} \; .
\hspace*{30mm}$$ Thus, we have shown (\ref{eq: abschaetzung der omegaabl der
greensfkt}) for $n=1$. We proceed inductively to conclude the proof.
\end{proof}
Note that the estimate (\ref{eq: abschaetzung der omegaabl der greensfkt}) remains
valid in the limit $0 \neq \omega \rightarrow 0$ for all $n$, because 
$$\lim_{\omega \rightarrow 0} \partial_\omega^n
\left( \frac{1}{\omega} \sin(\omega (u-v)) \right) = \left\{\begin{array}{cll}
\displaystyle (-1)^{n/2} \frac{1}{n+1} (u-v)^{n+1}&, & \textrm{if } n
\textrm{ even,} \vspace*{2mm}
\\ 
0 & , & \textrm{if } n \textrm{ odd.}
\end{array} \right.
$$

\begin{proof}[Proof of Lemma \ref{lemma: Entwicklung von acute phi,l=0}:]
First, remember that the solution $\grave{\phi}_\omega(u)$ is given by the
perturbation series $$\grave{\phi}_\omega (u) = \sum_{k=0}^\infty
\phi_\omega^{(k)} (u) \; ,$$ where the summands follow the iteration scheme
\begin{equation} \label{eq: iterationsschema fuer grave phi}\phi_\omega^{(0)} (u)  =
e^{ - i \omega u}\; , \; \phi_\omega^{(k+1)}(u) = 
 - \int_{u}^\infty \frac{1}{\omega} \sin(\omega(u-v))V_0(v)
\phi_\omega^{(k)}(v) \:dv \; , 
\end{equation}
with potential $ \displaystyle V_0(u) =  \left(1-
\frac{2M}{r(u)} \right) 
\frac{2M}{r(u)^3}$. According to Lemma \ref{lemma: asymptotische Entwicklung von
V_l}, this potential can be represented for large $u$ as $\displaystyle V_0(u) =
\frac{c_{30}}{u^3} +h(u)$, with $\displaystyle h(u) = \mathcal{O} \left(
\frac{\log u}{u^4}\right)$. Next, we split this iteration scheme up. To this end, we
define 
\begin{equation} \label{eq: definition phi tilde im beweis der entwicklung}
\tilde{\phi}^{(1)}_\omega (u) := -\int_{u}^\infty \frac{1}{\omega} \sin (\omega(u-v))
h(v) e^{-i \omega v} \: dv \; ,
\end{equation}
and analogously, 
\begin{equation} \label{eq: definition phi hut im beweis der entw.}
\hat{\phi}^{(1)}_\omega(u):= - \int_{u}^\infty \frac{1}{\omega} \sin (\omega(u-v))
 \frac{c_{30}}{v^3} e^{-i \omega v} \: dv \; .
\end{equation}
Thus, obviously $\phi_\omega^{(1)}(u) = \hat{\phi}_\omega^{(1)}(u) +
\tilde{\phi}_\omega^{(1)}(u)$. Now we iterate these two functions
$$ \tilde{\phi}_\omega^{(k+1)}(u) := - \int_{u}^\infty \frac{1}{\omega}
\sin(\omega(u-v))V_0(v) \tilde{\phi}_\omega^{(k)}(v) \:dv \; , \; k \geq 1 \; ,$$ 
analogously for $\hat{\phi}_\omega^{(k+1)}(u)$. Hence, we have the
formal decomposition 
\begin{equation} \label{eq: formal decomposition}
\grave{\phi}_\omega (u) = e^{-i \omega u} + \sum_{k=1}^\infty
\hat{\phi}_\omega^{(k)}(u) + \sum_{k=1}^\infty \tilde{\phi}_\omega^{(k)}(u) \;.
\end{equation}
Both series are well-defined. In order to show this, we use the bound
\begin{equation} \label{eq: alte Ungleichung fuer greensfkt}
 \Big| \frac{1}{\omega} \sin(\omega(u-v)) \Big| \: \leq \:\frac{4|v|}{1 +
|\omega v|} \; ,
\end{equation}
from Lemma \ref{lemma: abschaetzung der omegaabl der greensfkt} for real $\omega$
[Note that this
estimate is also
valid for the case $v\geq u>0$]. Hence, we get inductively the estimates
\begin{eqnarray*}
 \big| \hat{\phi}_\omega^{(k+1)} (u) \big| \leq \hat{R}_\omega (u)
\frac{P_\omega(u)^k}{k!} \; , \\
\big| \tilde{\phi}_\omega^{(k+1)} (u) \big| \leq \tilde{R}_\omega (u)
\frac{P_\omega(u)^k}{k!} \; ,  
\end{eqnarray*}
for all $k \geq 0$, where the functions $\hat{R}, \tilde{R}$ and $P$ are given by
\begin{eqnarray*}
\hat{R}_\omega (u) & := & \int_u^\infty  \frac{4 v}{1 + |\omega| v}
\Big|\frac{c_{30}}{v^3} \Big| \: dv \; ,\\
\tilde{R}_\omega (u) & := & \int_u^\infty  \frac{4 v}{1 + |\omega| v}
| h(v) | \: dv \; ,\\
P_\omega(u) & := & \int_u^\infty  \frac{4 v}{1 + |\omega| v}
| V_0 (v) | \: dv \; .
\end{eqnarray*}
Thus, the series $\sum \hat{\phi}_\omega^{(k)}(u)$ as well as $\sum
\tilde{\phi}_\omega^{(k)}(u)$ converge locally uniformly with respect to $u$ and
$\omega$. In the next step we show that, for fixed $u>0$, $\sum
\tilde{\phi}_\omega^{(k)}(u)$ is $C^1(\R)$ with respect to $\omega$. To this end, it
suffices to prove that each summand $ \tilde{\phi}_\omega^{(k)}, \; k\geq 1,$ is
$C^1$ and that the series $\sum \partial_\omega  \tilde{\phi}_\omega^{(k)}$
converges locally uniformly in $\omega$. Due to the estimates (\ref{eq: abschaetzung
der omegaabl der greensfkt}),(\ref{eq: alte Ungleichung fuer greensfkt}), we have the
inequality
\begin{eqnarray}
\Big| \partial_\omega \left[ \frac{1}{\omega} \sin (\omega(u-v)) h(v) e^{-i\omega
v} \right] \Big| \leq \hspace*{5cm} \nonumber \\ \leq \Big| \frac{12 \: v^2}{1 +|
\omega |v}
h(v) \Big| + \Big| \frac{4 v^2}{1 + |\omega| v} h(v)\Big| = \frac{16 v^2}{1 +
|\omega| v} |h(v)|\;. \label{eq: Abschaetzung Integralkern tilde phi 1}
\end{eqnarray}
Hence, the second term is an integrable bound, uniformly in $\omega$, for the first
derivative of the integrand. It follows that $\tilde{\phi}_\omega^{(1)}(u)$ is $C^1$
with respect to $\omega$, bounded by
\begin{eqnarray*}
\big| \partial_\omega \tilde{\phi}_\omega^{(1)}(u) \big| \leq \int_u^\infty \frac{16
v^2}{1 + |\omega| v} |h(v)| \:dv =: \tilde{R}_\omega^{(1)}(u) \; .
\end{eqnarray*}
Together with the estimate
\begin{eqnarray*}
\tilde{R}_\omega (u) \leq \frac{1}{4 u} \int_u^\infty \frac{16 v^2}{1 +|\omega| v}
|h(v)| \:dv \leq \frac{1}{u} \tilde{R}_\omega^{(1)}(u) \; ,
\end{eqnarray*}
one shows inductively that $\tilde{\phi}_\omega^{(k+1)} (u)$ is $C^1$ with respect
to $\omega$, bounded by
\begin{eqnarray*}
\big| \partial_\omega \tilde{\phi}_\omega^{(k+1)} (u) \big| \leq
\tilde{R}_\omega^{(1)} (u) \frac{(4 P_\omega(u))^k}{k!} \; .
\end{eqnarray*}
This yields that the sum $\sum \partial_\omega  \tilde{\phi}_\omega^{(k)}$ converges
locally uniformly in $\omega$. Hence, the sum  $\sum
\tilde{\phi}_\omega^{(k)}(u)$ is $C^1(\R)$ with respect to $\omega$. According to the
decomposition (\ref{eq: formal decomposition}), it remains to analyze the
$\omega-$dependence of $\sum \hat{\phi}_\omega^{(k)}(u)$. To this end, we compute the
first summand: 
\begin{eqnarray*}
\hat{\phi}_\omega^{(1)} (u) & = & \frac{1}{2 i \omega} \int_u^\infty \left( e^{-i
\omega(u-v)} - e^{i \omega (u-v)}\right) e^{-i \omega v} \frac{c_{30}}{v^3} \: dv
\\
& = & \frac{1}{2 i \omega}  e^{-i \omega u} \int_u^\infty \frac{c_{30}}{v^3} \: dv -
\frac{1}{2 i \omega} e^{i \omega u} \int_u^\infty  \frac{c_{30}}{v^3} e^{-2 i \omega
v} \: dv \; .
\end{eqnarray*}
Integrating the second term by parts, we obtain
\begin{eqnarray*}
& = & \frac{1}{2 i \omega} \left(  e^{-i \omega u} \frac{c_{30}}{2 u^2} -
 e^{i \omega u} \frac{c_{30}}{2 u^2} e^{-2 i \omega
u} + e^{i \omega u} \int_u^\infty  \frac{c_{30}}{-2v^2} (-2 i
\omega) e^{-2 i \omega v} \: dv \right) \\  
& = & e^{i \omega u} \int_u^\infty  \frac{c_{30}}{2v^2} \: e^{-2 i
\omega v} \: dv \; .
\end{eqnarray*}
The series expansion of Lemma \ref{lemma: zur Reihenentwicklung} in the limit
$\varepsilon \rightarrow 0$ yields
\begin{eqnarray} \label{eq: Entwicklung von hut phi 1}
\hat{\phi}_\omega^{(1)} (u) = \frac{c_{30}}{2} e^{i \omega u} \bigg\{ 2 i
\omega  \big( \log \left(2 i \omega u\right) +  c_0 \big) \hspace*{30mm} \nonumber 
\\ 
 - u^{-1} \sum \limits_{k=0,k\neq 1}^\infty \frac{(-1)^k }{(k-1) k!} \left(2 i
\omega u\right)^k  \bigg\} \; .
\end{eqnarray}
Intuitively, the only term which is not $C^1$ is the term involving $2 i \omega  
\log (2 i \omega u)$. More precisely, defining
\begin{eqnarray} \label{eq: Definition hut psi 1} \hat{\psi}_\omega^{(1)} (u) :=
\hat{\phi}_\omega^{(1)} (u) - 
c_{30} e^{i \omega u} i\omega  \log \left(2 i \omega u\right) \; ,
\end{eqnarray}
and iterating this by
\begin{equation*}
\hat{\psi}_\omega^{(k+1)} (u) := - \int_u^\infty \frac{1}{\omega} \sin(\omega(u-v))
V_0(v) \hat{\psi}_\omega^{(k)}(v) \: dv \; ,\; k \geq 1,
\end{equation*}
we show next that the sum $\sum \hat{\psi}_\omega^{(k)} (u)$ is $C^1$ with respect
to $\omega$. By definition this holds for the initial function
$\hat{\psi}^{(1)}_\omega (u)$. In order to prove this for the sum, we apply the same
method as above. To this end, we need good estimates for the initial functions
$\hat{\psi}_\omega^{(1)} (u)$ and $
\partial_\omega \hat{\psi}_\omega^{(1)} (u)$. 
Estimating the integral representation of $\hat{\phi}^{(1)}_\omega (u)$, we obtain
for arbitrary $u>0$ and $\omega \in \R$,
\begin{eqnarray*}
 \Big| \hat{\psi}_\omega^{(1)} (u) + c_{30} e^{i \omega u} 
i\omega \log \left(2 i \omega u\right) \Big| = \Big| \hat{\phi}_\omega^{(1)} (u)
\Big|   \leq \int_u^\infty \Big| \frac{c_{30}}{2 v^2} \Big| \: dv =
\frac{c_{30}}{2u} \; .
\end{eqnarray*} 
On the other hand, looking at the series
in (\ref{eq: Entwicklung von hut phi 1}), we obtain for all $u \leq
\frac{1}{| \omega|} $ the estimate
\begin{eqnarray}
\big| \hat{\psi}^{(1)}_\omega (u) \big| = \Big| \frac{c_{30}}{2} e^{i \omega u}
\Big\{ 2 i \omega  c_0  - \frac{1}{u} \sum \limits_{k=0,k\neq 1}^\infty \frac{(-1)^k
}{(k-1) k!} \left(2 i\omega u\right)^k  \Big \} \Big| \leq \frac{\tilde{c}}{u} \; ,
\label{eq: Abschaetzung hut psi 1 bereich leq 1/omega}
\end{eqnarray}
with a suitable constant $\tilde{c}$. Thus, we get for all $u>0$ and $ \omega \in \R$
the estimate
\begin{equation} \label{eq: abschaetzung hut psi 1}
\big| \hat{\psi}^{(1)}_\omega (u) \big| \leq \frac{c}{u} + c |\omega| |\log (2 i
\omega u) |  \: 1_{[\frac{1}{|\omega|} , \infty)} (u) \; ,
\end{equation} 
where $c$ is chosen suitably and $1_{.}(.)$ denotes the characteristic function.
In order to estimate the derivative $\partial_\omega \hat{\psi}^{(1)}_\omega (u)$, we
use in the domain $ u \geq \frac{1}{|\omega|}$, $ |\omega| \neq 0$, the following
bound for $\partial_\omega  \hat{\phi}^{(1)}_\omega (u)$ [see also (\ref{eq:
Abschaetzung Integralkern tilde phi 1})],
\begin{eqnarray*}
\nonumber \Big| \partial_\omega \left( \hat{\phi}^{(1)}_\omega (u)\right) \Big| 
& \leq &
\int_u^\infty \frac{16 v^2}{1 + |\omega| v} \Big| \frac{c_{30}}{v^3} \Big| \: dv
\\
& \leq & \frac{16}{|\omega|} \int_u^\infty \frac{|c_{30}|}{v^2} \: dv \leq \frac{16
c_{30}}{|\omega| u} \leq 16 \: c_{30} \; .
\end{eqnarray*}
Together with the analogon to estimate (\ref{eq: Abschaetzung hut psi 1 bereich
leq 1/omega}) in the region $u \leq \frac{1}{|\omega|}$, we obtain the bound
\begin{eqnarray} \label{eq: abschaetzung partial omega hut psi 1}
\big| \partial_\omega \hat{\psi}_\omega^{(1)} (u) \big| \leq \tilde{c} + \tilde{c}(1
+ u |\omega|) |\log(2i \omega u)| \: 1_{[\frac{1}{|\omega|}, \infty) } (u) \; , 
\end{eqnarray}
where $u>0,\omega \in \R$ and $\tilde{c}$ is an appropriate constant.
For reasons of simplicity, we choose $c = \tilde{c}$ such that both inequalities
(\ref{eq: abschaetzung hut psi 1}),(\ref{eq: abschaetzung partial omega hut psi 1})
hold. Using these inequalities, we show by induction, in the
same way as above, that $\hat{\psi}^{(k)}_\omega (u)$ is $C^1$ with
respect to $\omega$ and obeys the estimates
\begin{eqnarray}
\big| \hat{\psi}_\omega^{(k)} (u) \big| & \leq & \frac{c}{u}  \frac{P_\omega
(u)^{k-1}}{(k-1)!} + \frac{c}{u}\: r (|\omega|) \frac{P_\omega (u)^{k-2}}{(k-2)!}  \;
,  \label{eq: abschaetzung hut psi k}\\
\big| \partial_\omega \big( \hat{\psi}_\omega^{(k)} (u) \big) \big| & \leq &
c \frac{\big(4 P_\omega (u)\big)^{k-1}}{(k-1)!} + 5 c\: r (|\omega|)
\frac{\big(4 P_\omega (u)\big)^{k-2}}{(k-2)!} \; , \label{eq: abschaetzung partial 
omega hut psi k}
\end{eqnarray}
for all $k \geq 2,u>0$ and $\omega \in \R$, where $r$ is given by
\begin{equation*}
r(|\omega|) :=  \int _{\frac{1}{|\omega|}}^\infty \frac{4| \omega|
v^2}{1 +
|\omega| v} \: | V_0(v) | | \log(2 i \omega v)| \: dv \; .
\end{equation*}
Due to (\ref{eq: abschaetzung hut psi k}),(\ref{eq: abschaetzung partial omega hut
psi k}), the sums $\sum \hat{\psi}^{(k)}_\omega (u)$ and $\sum
\partial_\omega \hat{\psi}^{(k)}_\omega (u)$ converge locally uniformly in $\omega$.
Hence, we conclude that $\sum \hat{\psi}^{(k)}_\omega (u)$ is well defined and
continuously differentiable with respect to $ \omega$.

Thus, it remains to look at the term we get by the iteration of
\begin{equation*}
\vartheta^{(1)}_\omega (u) := c_{30} e^{i\omega u} i \omega \log(2 i \omega u) =
i c_{30} \: \omega \log(2 i \omega) e^{i\omega u} + i  c_{30} \: \omega\log(u) e^{i
\omega u} \; .
\end{equation*}
To this end, we split up the iteration, exactly as we did for the iteration of
$\phi^{(k)}_\omega (u)$, i.e. we define
\begin{eqnarray*}
\tilde{\vartheta}^{(2)}_\omega (u) & := & -\int_{u}^\infty \frac{1}{\omega} \sin
(\omega(u-v))
h(v) \vartheta^{(1)}_\omega (v) \: dv \; , \\
\hat{\vartheta}^{(2)}_\omega(u) & := &  - \int_{u}^\infty \frac{1}{\omega} \sin
(\omega(u-v))
 \frac{c_{30}}{v^3} \vartheta^{(1)}_\omega (v) \: dv \; ,
\end{eqnarray*}
and iterate these functions, 
$$ \tilde{\vartheta}_\omega^{(k+1)}(u) := - \int_{u}^\infty \frac{1}{\omega}
\sin(\omega(u-v))V_0(v) \tilde{\vartheta}_\omega^{(k)}(v) \:dv \; , \; k \geq 2 \;
,$$ 
analogously for $\hat{\vartheta}_\omega^{(k+1)}(u)$. Next, in exactly the
same way as for $\tilde{\phi}^{(k)}$, one sees that $$\sum_{k=2}^\infty
\tilde{\vartheta}_\omega^{(k)}(u) =2 i \omega \log (2
i \omega) \:f_1(\omega,u) +2 i \omega f_2 (\omega,u) \; ,$$ where $f_1(.,u)$ and
$f_2(.,u)$ are $C^1$ with respect to $\omega$. Finally, by an exact
calculation
\begin{eqnarray*}
\hat{\vartheta}^{(2)}_\omega (u) = i c_{30}^2 \omega \log(2 i \omega)e^{-i \omega u}
\int_u^\infty e^{2i \omega v} \frac{1}{2v^2} \: dv \hspace*{10mm}\\
+ i c_{30}^2 \omega e^{-i\omega u} \int_u^\infty e^{2 i \omega v}
\left(
\frac{1}{4v^2} + \frac{\log v}{2 v^2} \right) \: dv \; ,
\end{eqnarray*}
together with the series expansion of Lemma \ref{lemma: zur Reihenentwicklung} in
the limit $\varepsilon \rightarrow 0$ we obtain
\begin{eqnarray*}
\hat{\vartheta}^{(2)}_\omega (u) =  \frac{1}{4} i c_{30}^2 \: \omega (1 + 2 \log(2 i
\omega)) e^{-i \omega u} \hspace*{55mm} \\ \times \bigg[ (-2 i \omega) \big( \log
\left(-2 i \omega u\right) + c_0 \big)  
 - u^{-1} \sum \limits_{k=0,k\neq 1}^\infty \frac{(-1)^k }{(k-1) k!} \left(- 2 i
\omega u\right)^k  \bigg] \\
+ \frac{1}{2} i c_{30}^2 \: \omega e^{-i\omega u} \Bigg[ 
\sum \limits_{m=0}^1 \begin{pmatrix}
 1\\ m
\end{pmatrix} \log^{1-m} (u) \nonumber \bigg\{ (-2 i \omega ) \times
 \hspace*{25mm} \\ \left(
 \frac{(-1)^{m+2}}{m+1} \log^{m+1}\left(-2 i\omega u\right) + \sum
\limits_{k=0}^m c_k \log^k (-2 i\omega u)\right) \nonumber\\
 - u^{-1}\sum \limits_{k=0,k\neq p-1}^\infty \frac{(-1)^k (-1)^m \:
m!}{(k-1)^{m+1}
k!} (-2 i \omega u)^k  \bigg\} \Bigg] \; .
\end{eqnarray*}
Proceeding in the same way as for $\sum \hat{\psi}^{(k)}_\omega (u)$ [i.e. we 
omit the $\log \omega$-terms in the square brackets, and iterate these functions], we
again get terms of the form $$ 2i\omega \log (2 i \omega) \:f_3(\omega,u) +2i \omega
f_4 (\omega,u) $$ with continuously differentiable functions $f_3(.,u),f_4(.,u)$. So
after simplifications there remain terms of the form $$(2i \omega)^2
\log^s(2i\omega)
\log^r(-2i \omega) \log^t(u) e^{-i \omega u}\;.$$ These are obviously $C^1$ with
respect
to $\omega$ and so is their iteration, due to the fact that the additional
$\omega$-order yields directly integrable bounds for all $\omega$. This completes
the proof.
\end{proof}
Note that one can apply this idea of the proof to the case $l\geq 1$. This yields a similar result but also requires much more complex calculations according to the construction of the Jost solutions $\grave{\phi}_\omega$ [cf. \cite[Section 5]{Kronthaler}].

\section{The decay rate for spherical symmetric initial data}

According to the integral representation \ref{eq: Darstellung der Loesung l=0} the solution of the Cauchy problem for
compactly supported smooth
initial data $\Psi_0 \in C_0^\infty (\R)^2$ has the pointwise representation
\begin{eqnarray*}
 \Psi(t,u)=e^{-itH}
\Psi_0 (u) =\hspace*{70mm}\\ - \frac{1}{\pi}   \int_{\mathbb{R}} e^{-i \omega t}
\left(
\int_{\textrm{supp}\: \Psi_0} \mathrm{Im} \! \left(
\frac{\acute{\phi}_{\omega l}(u)
\grave{\phi}_{\omega l}(v)}{w(\acute{\phi}_{\omega l},\grave{\phi}_{\omega
l})}\right)
 \left(%
\begin{array}{lc}
  \omega & 1 \\
  \omega^2 & \omega \\
\end{array}%
\right)  \Psi_0(v) dv \right)  \: d\omega \; , 
\end{eqnarray*}
Our goal is now to use the Fourier transform (\ref{eq: Darstellung der Loesung
l=0}), in order to get detailed decay rates. To this end, we have to analyze the
integral
kernel, hence essentially
\begin{equation} \label{eq: essentieller part vom Integralkern} 
 \mathrm{Im} \! \left(\frac{\acute{\phi}_{\omega}(u)
\grave{\phi}_{\omega}(v)}{w(\acute{\phi}_{\omega },\grave{\phi}_{\omega
})}\right)\; .
\end{equation}

Since we already know that $\acute{\phi}_\omega$ is analytic on a 
neighborhood of the real line, it remains to understand $\grave{\phi}_\omega$ at
the point $\omega = 0$. To this end, we want to use an expansion as in Lemma
\ref{lemma: Entwicklung von acute phi,l=0}. The problem is that this expansion is not
sufficient for this purpose. Thus, we apply a similar method in order to gain

\begin{lemma} \label{lemma: bessere Entwicklung von gravephi for l=0}
For $l=0$, $\omega \in \R \setminus \{0\}$, $n\geq 3$ and fixed $u>0$, we get for the
fundamental solution $\grave{\phi}_\omega(u)$ the representation
\begin{equation}\label{eq: bessere Entwicklungvon gravephi for l=0}
\grave{\phi}_\omega (u) = e^{-i\omega u} + g_0(\omega,u)+ \sum_{i\geq j+k=1}^n
(2i\omega)^i \log^j(2i\omega)\log^k(-2i\omega) g_{ijk}(\omega,u) \; ,
\end{equation}
where the functions $g_0,g_{ijk} \in C^n(\R)$ with respect to $\omega$.
\end{lemma}

In order to prove this, we need the following lemma.
\begin{lemma} \label{lemma: Restiteration ist C3}
Let $u>0$, $n \in \N$ and $h \in C^\infty(\R_+)$ be a smooth function satisfying
$\int_u^\infty
v^{n+1} |h(v)| \:dv < \infty$. \\
Then:
\begin{enumerate}
\item $$ f^{(1)}_\omega (u) := - \int_u^\infty \frac{1}{\omega}
\sin(\omega(u-v)) h(v) e^{-i \omega v} \: dv$$ is $C^n(\R)$ with respect to
$\omega$.
\item For all $k \geq 1$
$$ f^{(k+1)}_\omega (u) := - \int_u^\infty \frac{1}{\omega} \sin(\omega
(u-v)) V_0(v) f^{(k)}_\omega (v) \:dv\; ,$$ are $C^n(\R)$ with respect to
$\omega$ and the series $ \sum_{k\geq1} \partial_\omega^m f^{(k)}_\omega (u)$,
$m \leq n$, converge locally uniformly.
\end{enumerate}
In particular, $\sum f^{(k)}_\omega (u)$ is $C^n(\R)$ with respect to
$\omega$.
\end{lemma}

\begin{proof}
This is shown in exactly the same way as the statement that the functions
$\tilde{\phi}^{(k)}_\omega$ in the proof of Lemma \ref{lemma: Entwicklung von acute
phi,l=0} as well as the series are $C^1$ with respect to $\omega$.
In order to show the differentiability up to the $n$-th order, we use the estimates
of
Lemma \ref{lemma: abschaetzung der omegaabl der greensfkt}.
\end{proof}

\begin{proof}[Proof of Lemma \ref{lemma: bessere Entwicklung von gravephi for l=0}]
Because of complex calculations we show this at first in the case $n=3$.
To this end, we split up the iteration scheme (\ref{eq: iterationsschema fuer grave
phi}) of the fundamental solutions in the following way. According to Lemma
\ref{lemma: asymptotische Entwicklung von V_l}, we can write the potential $V_0$ as
$$ V_0(v) = \sum_{p=3}^5 \sum_{q=0}^{p-3} c_{pq}\frac{\log^q(v)}{v^p} + r_6(v) \;
,$$ where $r_6$ is a smooth function for $v\geq u$ behaving asymptotically at
infinity as
$\mathcal{O}\left( \frac{\log^3(v)}{v^6}\right)$. Thus, defining 
$$\tilde{\phi}^{(1)}_\omega (u) := -\int_u^\infty \frac{1}{\omega}\sin(\omega(u-v))
r_6(v) e^{-i \omega v} \: dv$$ and for all $k\geq1$
$$ \tilde{\phi}^{(k+1)}_\omega (u):= - \int_u^\infty \frac{1}{\omega} \sin(\omega
(u-v)) V_0(v) \tilde{\phi}^{(k)}_\omega (v) \:dv\; ,$$ Lemma \ref{lemma:
Restiteration ist C3} yields that $\sum \tilde{\phi}^{(k)}_\omega \in C^3(\R)$ with
respect to $\omega$ and is a contribution to $g_0(\omega,u)$ in the statement of the
lemma.
Thus, we have to compute the remaining term
$$\hat{\phi}^{(1)}_\omega (u) := -\int_u^\infty \frac{1}{\omega}\sin(\omega(u-v))
\sum_{p=3}^5 \sum_{q=0}^{p-3} c_{pq}\frac{\log^q(v)}{v^p}
e^{-i \omega v} \: dv.$$
We do this essentially in the same way as we computed the terms $\hat{\phi}^{(1)},
\hat{\vartheta}^{(2)}$ in the proof of Lemma \ref{lemma: Entwicklung  von acute
phi,l=0}. We split up the $\sin(\omega(u-v))$ with Euler's formula and integrate
by parts and obtain
\begin{eqnarray} \label{eq: Integraldarstellung von phihut3}
= -e^{i \omega u} \int_u^\infty \left( \frac{c_{30}}{-2 v^2} + \sum_{p=3}^4
\sum_{q=0}^{p-2} \tilde{c}_{pq} \frac{\log^q v}{v^p} \right) e^{-2 i \omega v} \: dv
\; ,
\end{eqnarray}
where the coefficients $\tilde{c}_{pq}$ depend on the integral functions of the terms
$\log^r v / v^s$.
Now, we apply Lemma \ref{lemma: zur Reihenentwicklung} in the limit $\varepsilon
\searrow 0$ and get
\begin{eqnarray}
\nonumber  = & & e^{i \omega u} \frac{c_{30}}{2} \left\{ 2 i \omega \log(2i\omega u)
-
\frac{1}{u} \sum_{k=0}^\infty d_k(2 i \omega u)^k \right\} \\ \nonumber
& + & e^{ i \omega u} \frac{c_{41}}{3} \bigg\{ (2i \omega)^2 \left(
\frac{1}{4}\log^2(2 i \omega u) +  \log(2 i \omega u) (c -\frac{1}{2} \log u)
\right) \\
 \nonumber
&  & \hspace*{40mm}  +\ (c + c \log u) \frac{1}{u^2} \sum_{k=0}^\infty d_k (2i\omega
u)^k\bigg\} \\ \nonumber
& + &  e^{i\omega u} \bigg\{ (2i \omega)^3 \sum_{s+t=1}^3 c
\log^s(2i\omega u) \log^t u \\ \label{eq: expansion of phihut3} & & \hspace*{40mm}
+\sum_{m=0}^2 c \log^m u \frac{1}{u^3}  \sum_{k=0}^\infty d_k(2i \omega u)^k \bigg\}
\end{eqnarray}
with appropriate constants $c$ and $d_k$, which are of the form
$$  d_k = \frac{(-1)^k (-1)^m \: m!}{(k-p+1)^{m+1} k!} \; . \qquad \textrm{[cf. Lemma
\ref{lemma: zur Reihenentwicklung}]} $$
Since the series-terms are obviously $C^3(\R)$ with respect to $\omega$, this
expression of $\hat{\phi}_\omega^{(1)}(u)$ fits into the desired expansion (\ref{eq:
bessere Entwicklungvon gravephi for l=0}). In the next step we have to iterate
(\ref{eq: expansion of phihut3}). To this end, we treat each term in the curly
brackets separately. We show this exemplarily for the first term which we denote by
\begin{eqnarray} \label{eq: def alpha 1}
\alpha^{(1)} (u)  & := & e^{i \omega u} \frac{c_{30}}{2} \bigg\{ 2 i \omega
\log(2i\omega u) - \frac{1}{u} \sum_{k=0}^\infty d_k(2 i \omega u)^k \bigg\}  \\
\nonumber
& = & e^{i \omega u} \int_u^\infty \frac{c_{30}}{2 v^2} e^{-2i\omega v} \: dv \; .
\end{eqnarray}
In order to derive sufficient bounds for all $u>0$, we use different methods for the
regions $|\omega| u\geq 1$ and $|\omega| u<1$. First, let $u$ be such that $|\omega|
u\geq 1$, and by integrating by parts we get:
\begin{eqnarray}
\alpha^{(1)} (u)& = & e^{i \omega u} \int_u^\infty \frac{c_{30}}{2
v^2}\frac{1}{(-2 i \omega)^3} \partial_v^3 e^{-2i\omega v} \: dv \nonumber\\
& = & \frac{c_{30}}{4 i \omega} e^{- i \omega u}  \frac{1}{u^2} -\frac{c_{30}}{(2i
\omega)^2} e^{- i \omega u}  \frac{1}{u^3} + \frac{3c_{30}}{(2i
\omega)^3} e^{- i \omega u}  \frac{1}{u^4} \label{eq: part int bei alpha 1} \\ & &-
e^{ i \omega u} \int_u^\infty
\frac{12}{v^5} \frac{1}{(2 i \omega)^3} e^{- 2 i \omega v} \: dv \; . \nonumber
\end{eqnarray}

Using this expression and elementary integral estimates, we get for all $u>0$
satisfying $|\omega| u \geq 1$ the bounds
\begin{eqnarray}\nonumber \big| \alpha^{(1)} (u) \big| & \leq & c
\frac{1}{|\omega|u^2} \:, \\
\nonumber \big|\partial_\omega \alpha^{(1)} (u) \big| & \leq & c \frac{1}{|\omega|u}
\:, \\
\nonumber \big|\partial_\omega^2 \alpha^{(1)} (u) \big| & \leq & c \frac{1}{|\omega|}
\:, \\  \label{eq: Abschaetzung fuer alpha bereich u omega geq1}
 \big|\partial_\omega^3 \alpha^{(1)} (u) \big| & \leq & c \frac{u}{|\omega|} \:,
\end{eqnarray}
with suitable constants $c$. Moreover, comparing the infinite sum of (\ref{eq: def
alpha 1}) with the exponential function, one directly sees that it is $C^3$ with
respect to $\omega$. It satisfies for all $u>0$ with $|\omega| u <1$ the bounds
\begin{eqnarray}
\nonumber \Big|   \frac{1}{u} \sum_{k=0}^\infty d_k(2 i \omega u)^k
\Big| & \leq & \frac{c}{u} \\
\nonumber \bigg| \partial_\omega \bigg( \frac{1}{u} \sum_{k=0}^\infty d_k(2 i \omega
u)^k\bigg) \bigg| & \leq & c \\
\nonumber \bigg| \partial_\omega^2 \bigg( \frac{1}{u} \sum_{k=0}^\infty d_k(2 i
\omega u)^k \bigg)
\bigg| & \leq & c u\\ \label{eq: abschaetzungen fuer summe in alpha omega u leq 1}
\bigg| \partial_\omega^3 \bigg( \frac{1}{u} \sum_{k=0}^\infty d_k(2 i \omega u)^k
\bigg)
\bigg| & \leq & c u^2 \; .
\end{eqnarray}
Using (\ref{eq: Abschaetzung fuer alpha bereich u omega geq1}),(\ref{eq:
abschaetzungen fuer summe in alpha omega u leq 1}), we verify that
iterating the sum first with $r_6$ followed by the full iteration with the potential
$V_0$, we obtain a $C^3$-function. For the remaining term $c_{30}/2 e^{i \omega u} 2i
\omega \log(2i\omega u)$ we use the identity $\log(2 i \omega u) = \log (2 i \omega)
+ \log u$ together with Lemma \ref{lemma:
Restiteration ist C3} to show that the first iteration with $r_6$ followed by the
full iteration with the potential $V_0$ yields a term of the form $ 2i\omega
\log(2 i \omega) f_{110}(\omega,u) +\omega f_{100}(\omega,u)$, $f_{110},f_{100} \in
C^3(\R)$ with respect to $\omega$, fitting into the expansion (\ref{eq: bessere
Entwicklungvon
gravephi for l=0}).
Thus, it remains to compute the integral
\begin{equation*}
-\int_u^\infty \frac{1}{\omega} \sin(\omega(u-v)) \sum_{p=3}^5 \sum_{q=0}^{p-3}
c_{pq}\frac{\log^q(v)}{v^p} \alpha^{(1)} (v) \:dv \; .
\end{equation*}
We do this exemplarily for the term
\begin{equation} \label{eq: definition von beta}
\beta^{(2)} (u) := -\int_u^\infty \frac{1}{\omega} \sin(\omega(u-v)) 
\frac{c_{30}}{v^3} \alpha^{(1)} (v) \:dv \; .
\end{equation}
A complex calculation, using Lemma \ref{lemma: zur Reihenentwicklung} and Lemma
\ref{lemma: entwicklung des iterationsschemas der unendl summe}, which will be 
stated and proven afterwards, yields
\begin{eqnarray}
\nonumber \beta^{(2)}(u) =& \! \! \! & 2 i\omega \log (2 i\omega) e^{-i \omega u}
c \bigg\{ (-2i \omega) \log(-2i\omega u) - \frac{1}{u} \sum_{k=0}^\infty d_k (-2i
\omega u)^k \bigg\} \\
\nonumber & + \! \! \! & 2 i \omega e^{-i \omega u} \frac{c_{30}^2}{4} \bigg\{
(-2i\omega) \big(-\frac{1}{2} \log^2(-2i\omega u) \\ & & \nonumber \hspace*{4mm}+ 
\log(-2i\omega u)(c+  \log u)\big) 
- (c+c\log u) \frac{1}{u}\sum_{k=0}^\infty
d_k (-2i\omega u)^k \bigg\} \\
&  +\! \! \! & e^{-i \omega u} \bigg\{c (2i\omega)^2 \log(-2i\omega u) +  \label{eq:
berechnung von beta (2)}  \frac{1}{u^2} \sum_{k=0}^\infty d_k(-2i\omega u)^k
\bigg\}\;,
\end{eqnarray}
with suitable constants $c,d_k$. Hence $\beta^{(2)}(u)$ goes with (\ref{eq: bessere
Entwicklungvon gravephi for l=0}). So far, we cannot finish this scheme,
but if one has a close look, one sees that the most irregular term at
$\omega=0$, namely $2i\omega \log (2i\omega)$, now appears with a $1/u$ decay, while
the other irregularities appear with an additional $\omega$-power.
Furthermore,
due to the bounds (\ref{eq: Abschaetzung fuer alpha bereich u omega geq1}) together
with direct integral estimates, we obtain for all $u$ with $|\omega|u \geq 1$ the
bounds
\begin{eqnarray}
\nonumber \big| \beta^{(2)}(u) \big| & \leq & c\int_u^\infty \frac{v}{1 +|\omega|
v}\frac{1}{v^3} \frac{1}{v^2 |\omega|}\:dv \leq c \frac{1}{ u^4| \omega |} \\
\nonumber 
\big| \partial_\omega \beta^{(2)}(u)\big| & \leq & c\frac{1}{u^3 |\omega|} \\ 
\nonumber 
\big| \partial_\omega^2 \beta^{(2)}(u)\big| & \leq &c \frac{1}{u^2 |\omega|} \\
\label{eq: Abschaetzungen fuer beta(2) bereich omega u geq1}
\big| \partial_\omega^3 \beta^{(2)}(u)\big| & \leq & c\frac{1}{u |\omega|} \; .
\end{eqnarray}
Using in the region $|\omega|u <1$ for the sum-terms in (\ref{eq: berechnung von beta
(2)}) estimates analog to (\ref{eq: abschaetzungen fuer summe in alpha omega u leq
1}), we conclude in the same way as before, that iterating these first with $r_6$
followed by the full iteration with the potential $V_0$ and summing up, we obtain
$C^3-$terms.
We split up the remaining $\log$-terms by $\log(-2i \omega u) =  \log(-2i \omega) +
\log(u)$ and use Lemma \ref{lemma: Restiteration ist C3} to show that applying the
same
procedure yields terms that go with (\ref{eq: bessere Entwicklungvon gravephi for
l=0}). Hence, we have to analyze the integral
\begin{equation*}
-\int_u^\infty \frac{1}{\omega} \sin(\omega(u-v)) \sum_{p=3}^5 \sum_{q=0}^{p-3}
c_{pq}\frac{\log^q(v)}{v^p} \beta^{(2)} (v) \:dv \; ,
\end{equation*}
exemplarily we treat the term
\begin{equation} \label{eq: Definition gamma (3)}
\gamma^{(3)}(u) := -\int_u^\infty \frac{1}{\omega} \sin(\omega(u-v))
\frac{c_{30}}{v^3} \beta^{(2)} (v) \:dv \; .
\end{equation}
Computing this expression with the same methods one sees that the term with $2 i
\omega \log(2 i \omega)$ decays as $1/u^2$ and the $\omega^2 \log^s(\pm
2i\omega)$-terms decay
as $\log^t(u)/u$. With bounds analog to (\ref{eq: Abschaetzungen fuer beta(2)
bereich omega u geq1}),(\ref{eq: abschaetzungen fuer summe in alpha omega u leq 1})
the same procedure applies and yields terms that match with (\ref{eq: bessere
Entwicklungvon gravephi for l=0}). Once again it remains to analyze
\begin{equation*}
-\int_u^\infty \frac{1}{\omega} \sin(\omega(u-v)) \sum_{p=3}^5 \sum_{q=0}^{p-3}
c_{pq}\frac{\log^q(v)}{v^p} \gamma^{(3)} (v) \:dv \; ,
\end{equation*}
and exemplarily
\begin{equation*} 
-\int_u^\infty \frac{1}{\omega} \sin(\omega(u-v))\frac{c_{30}}{v^3} \gamma^{(3)}
(v) \:dv \; .
\end{equation*}
Calculating this, one checks that the $2 i \omega \log(2 i \omega)$-term
decays as $1/u^3$, the $\omega^2 \log^s(\pm 2i\omega)$-terms decay
as $\log^t(u)/u^2$ and the $\omega^3 \log^m(\pm 2i \omega)$-terms decay as
$\log^n(u)/u$. Applying this scheme two times more, all terms which are not $C^3$
with respect to $\omega$ decay at least as $\log^s(u)/u^3$. Subtracting these terms
from the full term, we obtain a $C^3$-term which is decaying at least as
$\log^s(u)/u^3$, according to estimates analog to (\ref{eq: Abschaetzungen fuer
beta(2) bereich omega u geq1}),(\ref{eq: abschaetzungen fuer summe in alpha omega u
leq 1}) and estimating $|\omega| $ by $1/u$ in the region $|\omega| u<1$. So Lemma
\ref{lemma: Restiteration ist C3} applies for the full iteration
with the potential $V_0$ and we get a $C^3$-term. Due to their decay, we are able to
iterate the subtracted $\log \omega$-terms also with the full potential $V_0$ and
get terms that match with (\ref{eq: bessere Entwicklungvon gravephi for l=0}).
Thus, the scheme can be stopped after finitely many calculations and the
lemma is proven for $n=3$. For $n \geq 4$ we split the potential in the way
$$ V_0(v) = \sum_{p=3}^{n+2} \sum_{q=0}^{p-3} c_{pq}\frac{\log^q(v)}{v^p} +
r_{n+3}(v) \; , $$ and proceed with the same calculations. In (\ref{eq: part int bei
alpha 1}) we have to integrate by parts up to the $n$-th order, in order to obtain as
analogon to estimate (\ref{eq: Abschaetzung fuer alpha bereich u omega geq1})
$$ \big|\partial_\omega^m \alpha^{(1)} (u) \big| \leq \frac{c}{|\omega|} u^{2-l} \;
, \quad m \leq n \; .$$ The next difference appears in the estimates (\ref{eq:
Abschaetzungen fuer beta(2) bereich omega u geq1}). These cannot be done for $n\geq
4$ by simple integral estimates as a matter of convergence. Thus, we have to subtract
from the result of the analog calculation to (\ref{eq: part int bei alpha 1}) for
$\alpha^{(1)}(u)$ the first $n-3$ exact terms of the form
$$ \frac{c}{\omega u^2}e^{-i \omega u} + ... +\frac{c}{\omega^{n-3} u^{n-2}} e^{-i
\omega u} =: \rho^{(1)}(u) \; ,$$
and get for $m \leq n$
\begin{eqnarray*}
\big| \partial_\omega^m \beta^{(2)}(u) \big|& \leq & \hspace*{2.7mm}\Big|
\partial_\omega^m \int_u^\infty \frac{1}{\omega} \sin(\omega(u-v))
\frac{c_{30}}{v^3} \left( \alpha^{(1)}(u) - \rho^{(1)} (u) \right) \Big| \\
& & + \Big| \partial_\omega^m \int_u^\infty \frac{1}{\omega} \sin(\omega(u-v))
\frac{c_{30}}{v^3} \rho^{(1)} (u)  \Big| \\ 
& \leq & \frac{c}{|\omega|} u^{m-4} \; ,
\end{eqnarray*} 
where for the first integral this can be done by elementary integral estimates, and
for the second integral we have to integrate the subtracted terms by parts, as we did
to obtain the estimates for $\alpha^{(1)}(u)$. Keeping these differences in mind, we
can conclude exactly in the same way as for $n=3$, which yields the claim for
arbitrary $n$.

\end{proof}
We now state the missing lemma.

\begin{lemma} \label{lemma: entwicklung des iterationsschemas der unendl summe}
Let $u>0$ and $\omega \in \R \setminus \{0\}$. For the calculation of the iteration
of the
infinite
sums that appear in the integration in Lemma \ref{lemma: zur Reihenentwicklung} with
an arbitrary part of the potential, $\log^q u/u^p$, cf. Lemma \ref{lemma:
asymptotische
Entwicklung von V_l}, we obtain the identity
\begin{eqnarray}
 -  & \hspace*{-2mm}\displaystyle\int_u^\infty \hspace*{-2mm} & \frac{1}{\omega} \sin
(\omega (u-v)) \frac{\log^q v}{v^p}
\frac{\log^s
v}{v^t} e^{\pm i \omega v} \sum_{k=0}^\infty d_k (\pm 2 i \omega v)^k \: dv \nonumber
\\
&  =& \frac{1}{u^{p+t-2}} e^{\mp i\omega u} \sum_{l=0}^{s+q} c\log^{l} (u)
\sum_{k=0}^\infty
d_{kl} (\mp 2 i \omega u)^k \\ \label{eq: entwicklung des iterationsschemas der
unendl
summe}& & + (2i \omega)^{p+t-2} e^{\mp i \omega u} \sum_{m=0}^{s+q} \sum_{r=1}^{m+1}
c
\log^r (\mp 2 i \omega u) \log^{s+q-m} (u) \; ,
\end{eqnarray}
for suitable constants $d_{kl},c$.
\end{lemma}

\begin{proof}
Let us denote $m=q+s \geq 0$ and $n=p+t \geq 4$. In order to compute the integral on
the left hand side in the lemma, we insert a convergence generating factor
\begin{eqnarray}
-  & \hspace*{-2mm}\displaystyle\int_u^\infty \hspace*{-2mm} & \frac{1}{\omega} \sin
(\omega (u-v)) \frac{\log^m
v}{v^n} e^{\pm i \omega v} \sum_{k=0}^\infty d_k (\pm 2 i \omega v)^k \: dv
\nonumber \\
&=& \lim_{\varepsilon \searrow 0} \int_u^\infty e^{- \varepsilon v}
\: \frac{1}{\omega} \sin (\omega (v-u))
 \frac{\log^m v}{v^n} e^{\pm i \omega v} \sum_{k=0}^\infty d_k
(\pm 2 i \omega v)^k \: dv . \hspace*{10mm} \label{eq: mit konvergenzerzeugendem
Faktor}
\end{eqnarray}
In the next step we interchange the integral and the infinite sum. This can be done
for any $\varepsilon > 2|\omega|$ by a dominating convergence argument, if one
estimates the modulus of the sum very roughly by $\exp(2|\omega| v)$. Thus, the two
expressions 
\begin{eqnarray*}
 \int_u^\infty \frac{1}{\omega} \sin (\omega (v-u))
e^{- \varepsilon v} \frac{\log^m v}{v^n} e^{\pm i \omega v} \sum_{k=0}^\infty d_k
(\pm 2 i \omega v)^k \: dv \\ \textrm{and} \hspace*{108mm} \\
 \sum_{k=0}^\infty d_k (\pm 2 i \omega)^k
\int_u^\infty \frac{1}{\omega} \sin (\omega (v-u))
e^{- \varepsilon v} \frac{\log^m v}{v^n} e^{\pm i \omega v} v^k \: dv 
\end{eqnarray*}
coincide for any $\varepsilon > 2 |\omega|$. Moreover, both
expressions are analytic in $\varepsilon$ for $ \Re
\varepsilon >0$. So by the identity theorem for analytic functions both expressions
coincide for any $\varepsilon >0$. So (\ref{eq: mit konvergenzerzeugendem Faktor}) is
equal to 
\begin{eqnarray*}
\lim_{\varepsilon \searrow 0} \sum_{k=0}^\infty d_k (\pm 2 i \omega)^k
\int_u^\infty \frac{1}{\omega} \sin (\omega (v-u))
e^{- \varepsilon v} \frac{\log^m v}{v^{n-k}} e^{\pm i \omega v} \: dv \; .
\end{eqnarray*}
Once again we rewrite $\sin(\omega(v-u))$ with Eulers formula and integrate by parts
[note that one has to be careful with the $\varepsilon$-terms that are generated by
this integration by parts, but in the limit $\varepsilon \searrow 0$ they vanish] to
obtain
\begin{eqnarray*}
= \lim_{\varepsilon \searrow 0} \sum_{k=0}^\infty d_k (\pm 2 i \omega)^k e^{\mp i
\omega u} \int_u^\infty  e^{\pm 2 i \omega v - \varepsilon v}
\sum_{l=0}^m c \frac{\log^l v}{v^{n-k-1}} \: dv \; ,
\end{eqnarray*}
with suitable constants $c$ arising from the integral function of
$\log^mv/v^{n-k}$. Now we apply Lemma \ref{lemma: zur Reihenentwicklung}, Lemma
\ref{lemma: zur Reihenentwicklung mit p<0}, take the limit $\varepsilon
\searrow 0$ and get
\begin{eqnarray*}
= \sum_{l=0}^m \sum_{k=0}^\infty d_k (\pm 2 i \omega)^k e^{\mp i
\omega u} \sum_{i=0}^l c\begin{pmatrix}
 l\\ i \end{pmatrix} \log^{l-i} (u) \times \hspace*{35mm}\\
\bigg\{ (\mp 2 i \omega)^{n-k-2} \sum_{j=0}^{i+1} c \log^j[\mp 2i \omega u] -
\frac{1}{u^{n-k-2}} \sum_{r=0,r\neq n-k-2}^\infty d_r (\mp2 i \omega u)^r\bigg\}.
\end{eqnarray*}
We reorder the two infinite sums to one infinite sum, which can
be done because of the structure of the coefficients $d_k,d_r$ of the exponential
integral that lets us compare the new coefficients to the coefficients of
the exponential series, and get the expression
(\ref{eq: entwicklung des iterationsschemas der unendl summe}).
\end{proof}

Next, we need a similar expansion for the derivative $\grave{\phi}'_\omega (u)$.
\begin{lemma} \label{lemma: Entwicklung von phi'}
For $l=0$, $\omega \in \R \setminus \{0\}$, $n \geq 3$ and fixed $u>0$, the first
$u$-derivative
of $\grave{\phi}_\omega (u)$ satisfies the expansion
\begin{equation} \label{eq: Entwicklung von phi'}
\grave{\phi}'_\omega (u) =  -i \omega e^{-i \omega u} + h_0(\omega,u) +
\sum_{i\geq j+k=1}^n (2i\omega)^i \log^j(2 i \omega) \log^k(-2i \omega)
h_{ijk}(\omega,u)
\; ,
\end{equation}
where the functions $h_0,h_{ijk} \in C^n(\R)$ with respect to $\omega$.
\end{lemma}

\begin{proof}
In order to prove this, we use the fact that $\grave{\phi}'_\omega (u)$ satisfies
for $u>0$  an integral equation analog to (\ref{Jost equation bound cond -unendlich})
\begin{equation*}
\grave{\phi}'_\omega (u) = -i \omega e^{- i \omega u} - \int_u^\infty \cos(\omega
(u-v)) V_0(v) \grave{\phi}_\omega (v) \: dv \; .
\end{equation*} 
We estimate $\cos(\omega(u-v))$ and its $\omega-$derivatives for real $\omega$ and
$v\geq u>0$ by
\begin{equation} \label{eq: Abschaetzung fuer cos omega u-v}
 \big|\partial_\omega ^n \cos(\omega (u-v)) \big| \leq (v-u)^n \leq
(2v)^n \; ,\quad n \in \N_0 \; .
\end{equation}
Thus, using this estimate for $n=0$ together with the iteration scheme (\ref{eq:
iterationsschema fuer grave phi})
for $\grave{\phi}_\omega (u)$, we obtain a well defined iteration scheme for the
$u$-derivative:
\begin{eqnarray}
\grave{\phi}'_\omega (u) &=& \sum_{k=0}^\infty \psi_\omega^{(k)}(u) \;
, \quad \textrm{where}\nonumber \\
\psi^{(0)}_\omega (u) &=& -i \omega e^{-i \omega u} = \left(\phi^{(0)}_\omega\right)'
(u) \; , \label{eq: Iterationsschema fuer phi'}
\\
\psi_\omega^{(k+1)}(u) &=& - \int_u^\infty \cos(\omega (u-v)) V_0(v)
\phi_\omega^{(k)} (v) \: dv  = \left(\phi^{(k+1)}_\omega\right)'(u)\; , 
 \nonumber
\end{eqnarray}
with $k\geq0$.
Due to this iteration scheme together with the estimates (\ref{eq: Abschaetzung fuer
cos omega u-v}) that replace the bounds (\ref{eq: abschaetzung der omegaabl der
greensfkt}) and the identity $\cos(\omega (u-v)) = 1/2 (e^{i\omega (u-v)} + e^{i
\omega (v-u)})$, we can use the decompositions of the $\phi_\omega^{(k)}$, which we
have made in the proof of Lemma \ref{lemma: bessere Entwicklung von gravephi for
l=0}. In particular, we apply the procedure of this proof, in order to show the
claim.
\end{proof}

Now, we use the expansions (\ref{eq: bessere Entwicklungvon gravephi for
l=0}),(\ref{eq: Entwicklung von phi'}), in order to analyze
the $\omega$-depen\-dence of the essential part of the integral kernel 
$$\mathrm{Im} \! \left(\frac{\acute{\phi}_{\omega}(u)
\grave{\phi}_{\omega}(v)}{w(\acute{\phi}_{\omega },\grave{\phi}_{\omega
})}\right)\; . $$ 
At this stage, it is enough to set $n=4$ in (\ref{eq: bessere
Entwicklungvon gravephi for l=0}),(\ref{eq: Entwicklung von phi'}) for our purposes.
Looking at the integral representation (\ref{eq: Darstellung der
Loesung l=0}) of the solution, we see that $u \in \R$ is fixed while $v \in
\R$ varies in a compact set, the support of our initial data $\Psi_0$. Due to the
Picard-Lindel{\"o}f theorem and the analytical dependence in $\omega$ of the
Schr{\"o}dinger equation from the coefficients, the expansions (\ref{eq: bessere
Entwicklungvon gravephi for l=0}),(\ref{eq: Entwicklung von phi'}) extend to any $u$,
and $v$, respectively, on compact sets. Moreover, the following properties follow
directly by the construction of the expansions.

\begin{corollary} \label{corollary: Beziehung g_ijk g_0}
For $4 \geq i=j+k \geq 1$ the function $g_{ijk},h_{ijk}$ can be constructed such that
they obey the equalities
\begin{equation} \begin{array}{rclll}
g_{ijk} (\omega, u) + o(\omega^\kappa)& = & c_{ijk} \left( e^{-i\omega u}
+g_0(\omega,u)
\right) 
\; & \mathrm{and}  \\ 
h_{ijk}(\omega,u)+ o(\omega^\kappa) & = & c_{ijk} \,
h_0(\omega,u)  \; , & & \textrm{ for } i \; \mathrm{even} \;
\\
g_{ijk} (\omega, u) + o(\omega^\kappa)& =  &c_{ijk} ( e^{i\omega u}
+\overline{g_0(\omega,u)}
)  \;& \mathrm{and}  \\ 
h_{ijk}(\omega,u) + o(\omega^\kappa) & =& c_{ijk} \,
\overline{h_0(\omega,u)}  \; ,& & \textrm{ for } i  \; \mathrm{odd} . 
\end{array} \label{eq: Beziehung gijk g0 hijk h0}
\end{equation}
where $\kappa$ is an arbitrary integer and the $c_{ijk}$ are real constants, in
particular not depending on $u$.
\end{corollary}

\begin{proof}
We show this exemplarily for the first terms $g_{110},h_{110}$.
In this situation, (\ref{eq: Beziehung gijk g0 hijk h0}) holds because the first
term,
where $(2i\omega) \log(2i \omega)$ appears, appears with $c_{30}/2 \; e^{i \omega u}$
and there are no other terms with this $\omega$-dependence except the terms
that are generated by this [cf. the calculations (\ref{eq: expansion of
phihut3}),(\ref{eq: berechnung von beta (2)})]. Thus, $g_{110} (\omega,u)$ is
generated by $e^{i \omega u}$, which is just the complex conjugate of $e^{-i\omega
u}$,
and this behavior is kept by the iteration scheme. So any $C^4$-term that is 
generated is the complex conjugate of a corresponding term of $g_0$. This is valid,
until one finishes the iteration scheme with the arguments at the end of the proof of
Lemma \ref{lemma: bessere Entwicklung von gravephi for l=0}, by what the
$o(\omega^\kappa)$-term arises. Since one can do arbitrary many calculations and in
each
iteration at least a $\pm 2i \omega \log(\pm 2i \omega)$ is generated, the $\kappa$
can be
chosen arbitrary. Moreover, looking at the iteration scheme (\ref{eq:
Iterationsschema fuer phi'}), the equalities for $h_{110}(\omega,u)$ are a
consequence
of the arguments for $g_{110}(\omega,u)$, because of the fact that by the
calculations concerning this scheme no additional highest order $\log$-terms, i.e.
$i=j+k$, are
generated.
\end{proof}

In the following assume that $\kappa=5$. We expand the functions $g_{ijk}(\omega,u)$
and
$h_{ijk}(\omega,u)$ in their Taylor polynom with respect to $\omega$ at $\omega = 0$
up to the fourth order:
\begin{eqnarray*}
g_{ijk} (\omega, u)= \sum_{m=0}^4 \frac{1}{m!} \partial_\omega^m
g_{ijk}\big|_{(0,u)} \omega^m + r_{ijk}(\omega,u) \; , \\
h_{ijk} (\omega, u)= \sum_{m=0}^4 \frac{1}{m!} \partial_\omega^m
h_{ijk}\big|_{(0,u)} \omega^m + q_{ijk}(\omega,u) \; ,
\end{eqnarray*}
where the remaining terms $r_{ijk}(\omega,u),q_{ijk}(\omega,u) \in C^4(\R)$ behave
for
small $\omega$ as $o(|\omega|^4)$. Note that, due to this fact, any 
logarithmic irregularity multiplied with $r_{ijk},q_{ijk}$ yields a $C^4$-term with
respect to $\omega$. Moreover, we expand for fixed $u$ the fundamental solution
$\acute{\phi}_\omega (u)$ and its $u-$derivative $\acute{\phi}'_\omega (u)$
\begin{eqnarray*}
\acute{\phi}_\omega (u) = \sum_{k=0}^\infty c_k(u) \omega^k \; ,
\quad \acute{\phi}'_\omega (u) = \sum_{k=0}^\infty d_k(u) \omega^k \; ,
\end{eqnarray*}
which exist, because these are analytic in $\omega$ for fixed $u$.
Since the fundamental solutions $\acute{\phi},\grave{\phi}$ are real for $\omega =0$,
the coefficients $g_{0}(0,u),h_{0}(0,u),c_0(u)$ and $d_0(u)$ are real for all
$u\in \R$. Using all these properties, we expand 
\begin{equation} \label{eq: essentieller teil des integralkerns ohne im}
\frac{\acute{\phi}_\omega(u) \grave{\phi}_\omega (v)}{w(\acute{\phi},
\grave{\phi})} \; ,
\end{equation}
with the ansatz of a geometrical series with respect to $\omega$. Note that,
according
to a result in \cite[Section 6]{Kronthaler} the Wronskian does
not vanish for $\omega = 0$. By a
straightforward calculation it is shown that, essentially using (\ref{eq: Beziehung
gijk g0
hijk h0}), the terms with the highest logarithmic order, i.e. $(2 i \omega)^i
\log(2i\omega)^j\log(-2i\omega)^k, i=j+k,$ vanish. Thus, we have to pick out the
terms $(2i\omega)^2 \log(2i \omega)^j \log(-2 i \omega)^k $ with $j+k=1$, in
order to get the lowest regularity. Looking at the calculations (\ref{eq: expansion
of phihut3}) and (\ref{eq: berechnung von beta (2)}) [Note that these are the only
possible terms, where a term with this irregularity appears the first time, according
to our construction. The others are just a consequence out of these and hence a
contribution to
functions $g_{2jk}$], the desired terms appear in $\grave{\phi}$ the first time as
\begin{equation} \label{eq: wichtigster Term der Entwicklung}
e^{-i \omega u} \left((2i\omega)^2 \log(2i \omega) (c+c\log u)  -
c(2i\omega)^2 \log(-2i\omega) \right) \; ,
\end{equation}
where a $(2i \omega)^2 \log(2i \omega) \log u$ shows up in the first
line of (\ref{eq: berechnung von beta (2)}), if one separates $\log(-2i \omega u) =
\log(-2i \omega) + \log u$. All other such terms appearing in the second line of
(\ref{eq: berechnung von beta (2)}) as well as in the second line of (\ref{eq:
expansion of phihut3}) vanish because of their coefficients. Applying the same
arguments as before, it follows that $   g_{201}(\omega,u)+o (\omega^\kappa)=
c(e^{-i\omega u} + g_0(\omega,u))$ and $  h_{201}(\omega,u) +o(\omega^\kappa)= c
h_0(\omega,u)$. Hence, the terms with
$(2i\omega)^2 \log(-2 i \omega)$ cancel in the $\omega$-expansion of (\ref{eq:
essentieller teil des integralkerns ohne im}). Because of the additional $\log
u$-term, we get 
$$   g_{210}(\omega,u)+o(\omega^\kappa) = c_1( e^{-i\omega u} +  g_0(\omega,u)) + 
c_2
(e^{-i \omega
u}\log u+ g(\omega,u) )\; ,$$
and
$$ h_{210}(\omega,u) +o(\omega^\kappa) = c_1 h_0(\omega,u) + c_2 h(\omega,u) +
\frac{1}{4}
c_{30}
e^{i \omega u} \; ,$$
with appropriate real constants $c_1,c_2$, where the last term appears by a direct
calculation of $\psi^{(1)} (u)$ with the part $c_{30}/ v^3$ of the potential
$V_0(v)$.
Furthermore, $g(\omega,u),h(\omega,u)$ are $C^4$-functions with respect to
$\omega$, where $g(\omega,u)$ is generated by the iteration of $e^{-i \omega u} \log
u$ and $h(\omega,u)$ the consequence out of this in (\ref{eq: Iterationsschema fuer
phi'}). One directly verifies that $g(0,u),h(0,u)$ are real, in general
non-vanishing. Putting all
these informations together, one sees that there appears a term with $(2i\omega)^2
\log(2 i \omega)$ in the $\omega$-expansion of (\ref{eq: essentieller teil des
integralkerns ohne im}), which is generated on the one hand by the $g(0,u),h(0,u)$,
and on the other hand by the $2i\omega \log(2i\omega)$-part multiplied with the
$ \omega$-contribution of first order of $\acute{\phi},\acute{\phi}'$. This
represents the
part with the highest irregularity with respect to $\omega$. Moreover, the related
coefficients are purely \textit{real}, depending on $u,v$ and in general
non-vanishing.
Using the identity
$$ \log(2i\omega) = i \frac{\pi}{2} \sign(\omega) + \log(2 |\omega|) \; ,$$
and taking the imaginary part of (\ref{eq: essentieller teil des integralkerns ohne
im}), which is just the essential part of our integral kernel, we obtain as the
lowest regular $\omega$-term in the expansion of (\ref{eq: essentieller part vom
Integralkern}) at $\omega = 0$
\begin{equation} \label{eq: kleinste regularitaet vom integralkern}
c_0(u)g_{20}(v) \: \omega^2 \sign(\omega) \; ,
\end{equation}
where the function $g_{20}(v)$ arises out of the foregoing calculation. 
The symmetry of (\ref{eq: essentieller part vom Integralkern}) with respect
to $u,v$ yields immediately $g_{20}(v) = k c_0(v)$ with an appropriate constant $k
\neq 0$.

\bigskip

In the next step we want to use (\ref{eq: kleinste regularitaet vom integralkern}),
in order to derive the decay of the solution $\Psi (t,u)$ given by (\ref{eq:
Darstellung der Loesung l=0}). To this end, first we have to analyze the behavior of
the $\omega$-derivatives of the
integrand up to the fourth order for large $|\omega|$.

\begin{lemma} \label{lemma: abfall der ome abln des integranden}
For $u \in \R$ and compactly supported smooth initial data $\Psi_0 \in
C^\infty_0(\R)^2$ of the Cauchy problem, the $\omega$-derivatives of the integrand
in the
integral representation (\ref{eq: Darstellung der Loesung l=0}) 
\begin{equation} \label{eq: integrand in der integraldarstellung}
\partial_\omega^m\left(
\int_{\mathrm{supp}\: \Psi_0} \mathrm{Im} \! \left(
\frac{\acute{\phi}_{\omega}(u)
\grave{\phi}_{\omega}(v)}{w(\acute{\phi}_{\omega },\grave{\phi}_{\omega
})}\right)
 \left(%
\begin{array}{lc}
  \omega & 1 \\
  \omega^2 & \omega \\
\end{array}%
\right)  \Psi_0(v) dv \right) \; , \quad m\in \{0,...,4\} \; ,
\end{equation}
have arbitrary polynomial decay in $\omega$ for $|\omega| \rightarrow \infty$.
\end{lemma}

\begin{proof}
We proceed essentially as in the proof of \cite[Theorem 6.5]{Kronthaler}. To this end, we have to investigate the behavior of 
$\acute{\phi}_\omega (u),\grave{\phi}_\omega (v)$ in $\omega$ for $u \in \R$ fixed
and $v$ in the compact set $\mathrm{supp} \Psi_0$. We start with
$\grave{\phi}_\omega$.
We assume that
$|\omega| \geq 1$ and $u_0 \in \R$ is arbitrary. Obviously, we find for
any $v \geq u \geq u_0$ and $m \in \{0,...,4\}$ a constant $C_1(u_0)$ such that 
\begin{equation} \label{eq: elementare abschaetzung der omega-ableitungen der
greensfktn}
\Big| \partial_\omega^m \left[ \frac{1}{\omega} \sin(\omega(u-v)) \right] \Big| \leq
 \frac{1}{|\omega|} C_1(u_0) (1 +|v|)^m \; .
\end{equation}
Furthermore, splitting the potential as $$V_0(u) = \sum_{p=3}^5 \sum_{q=0}^{p-3}
c_{pq}\frac{\log^q(v)}{v^p} + r_6(v)$$ and following an analog calculation as
in (\ref{eq: part int bei alpha 1}), we obtain for the $\omega$-deriva\-tives of the
first iteration $\phi^{(1)}_\omega (u)$ for all $u \geq 1$ and $m \in\{0,...,4\}$
the estimate
\begin{equation} \label{eq: absch omega abl 1.Iteration}
\Big| \partial_\omega^m \phi_\omega^{(1)} (u) \Big| \leq \frac{1}{|\omega|} C_2
u^{m-2} \; ,
\end{equation}
with an appropriate constant $C_2$. [Note that this is just an analogue to the
estimate (\ref{eq:
Abschaetzung fuer alpha bereich u omega geq1}).] For all $u<1$ and $m \in\{0,...,4\}$
we get
\begin{eqnarray*}
\Big| \partial_\omega^m \phi_\omega^{(1)} (u) \Big| & \leq & \Big| \partial_\omega^m
\int_u^1 \frac{1}{\omega} \sin(\omega(u-v)) V_0(v) e^{-i \omega v} \: dv \Big| \\ 
 & & + \Big| \partial_\omega^m \int_1^\infty \frac{1}{\omega} \sin(\omega(u-v)) V_0(v
) e^{-i \omega v} \: dv\Big| \\
& \leq &  \frac{1}{|\omega|} f(m,u) + C_3 \frac{1}{|\omega|} \sum_{k=0}^m |u|^k
\; ,
\end{eqnarray*}
where $f$ is a continuous function with respect to $u$ and the second term arises
by the same method as we used for the estimate (\ref{eq: absch omega abl
1.Iteration}). Defining $C_4$ by
$$ C_4 := \max_{m \in \{0,...,4\}}  \max_{u \in [u_0,1]} \left\{ \left(f(m,
u) + C_3 \frac{1}{|\omega|} \sum_{k=0}^m |u|^k\right) (1+|u|)^{2-m} \right\} \; ,$$
and $C_5 :=\max(C_2,C_4)$, we obtain for all $u \geq u_0$ and $m\in
\{0,...,4\}$ the bound
\begin{equation} \label{eq: Schranke omega abl 1.Iteration alle u}
\Big| \partial_\omega^m \phi_\omega^{(1)} (u) \Big| \leq \frac{1}{|\omega|} C_5
(1+|u|)^{m-2} \; .
\end{equation}
In order to estimate the derivatives of the second iteration $\phi_\omega^{(2)} (u)$
up to the fourth order, we subtract the first exact term out of the integration by
parts in (\ref{eq: absch omega abl 1.Iteration}), $\displaystyle
\frac{c_{30}}{4 i \omega u^2} e^{-i \omega u}$, from the first
iteration $\phi_\omega ^{(1)} (u)$ and obtain for $u \geq1$ and $m\leq 4$ the
bounds
\begin{equation} \label{eq: Fall l=4 absch iteration - exakter term}
 \Big| \partial_\omega^m \left( \phi^{(1)}_\omega (u) - \frac{c_{30}}{4 i
\omega u^2} e^{-i \omega u}\right) \Big| \leq \frac{1}{|\omega|} C u^{m-3}.
\end{equation}
Thus, in order to estimate the $\omega$-derivatives of the second iteration:
\begin{eqnarray*}
\nonumber \Big| \partial_\omega^m  \phi^{(2)}_\omega (u) ) \Big| & \hspace*{-2mm}
\leq \hspace*{-2mm}
& \Big| \partial_\omega^m \int_u^\infty \frac{1}{\omega} \sin(\omega(u-v)) V_0(v)
\left( \phi^{(1)}_\omega (v) - \frac{c_{30}}{4 i
\omega u^2} e^{-i \omega v}\right) \: dv \Big|  \\
& & + \Big| \partial_\omega^m \int_u^\infty \frac{1}{\omega} \sin(\omega(u-v)) V_0(v)
 \frac{c_{30}}{4 i
\omega u^2} e^{-i \omega v} \:dv\Big| \; . 
\end{eqnarray*}
Using the estimates (\ref{eq: Fall l=4 absch iteration - exakter term}),(\ref{eq:
elementare abschaetzung der omega-ableitungen der greensfktn}), and once again the
method of splitting up the potential and integrating by parts for the second
integral, we get for $u\geq1$ and $m\leq4$ the bounds
\begin{equation}
\nonumber \Big| \partial_\omega^m  \phi^{(2)}_\omega (u) ) \Big| \leq
\frac{1}{|\omega|} C u^{m-4},
\end{equation}
and thus, following the foregoing arguments for all $u \geq u_0$ (after possibly
enlarging $C_5$) the estimates
\begin{equation} \label{eq: Schranke omega abl 2.Iteration alle u}
\Big| \partial_\omega^m  \phi^{(2)}_\omega (u) ) \Big| \leq
\frac{1}{|\omega|} C_5 (1 + |u|)^{m-4} \; .
\end{equation}

Using (\ref{eq: elementare abschaetzung der omega-ableitungen der greensfktn}) and
(\ref{eq: Schranke omega abl 2.Iteration alle u}), we obtain for the
$\omega$-derivatives of the third iteration for all $u\geq u_0$
\begin{eqnarray}
\nonumber
\Big| \partial_\omega^m \phi_\omega^{(3)} (u) \Big|&\hspace*{-2mm} \leq
\hspace*{-2mm} & \Big| \sum_{k=0}^m
\begin{pmatrix} m \\ k \end{pmatrix}  \int_u^\infty \partial_\omega^{m-k} \left(
\frac{1}{\omega} \sin(\omega(v-u))\right)V_0(v) \partial_\omega^k
\phi_\omega^{(2)} (v) \: dv \Big| \\
& \hspace*{-2mm} \leq \hspace*{-2mm} & 16 C_1(u_0) C_5 \frac{1}{|\omega|}
\int_u^\infty (1+|v|)^{m-4}
\frac{1}{|\omega|}V_0(v) \:dv \label{eq: 2.Iteration Induktionsanfang} \; .
\end{eqnarray}
Note that interchanging the integral and the $\omega$-derviatives is permitted,
because the $\omega$-derivatives of the integrand are integrable due to the
estimates (\ref{eq: elementare abschaetzung der omega-ableitungen der
greensfktn}),(\ref{eq: Schranke omega abl 2.Iteration alle u}) and the $1/v^3$-decay
of the
potential $V_0(v)$. We show by induction
in $n$ for all $u\geq u_0$ the inequality
\begin{equation} \label{eq: Induktion der absch der ome abln}
\Big| \partial_\omega^m \phi_\omega^{(n)} (u) \Big| \leq 16 C_1(u_0) C_5
\frac{1}{|\omega|} Q_\omega(m,u) \frac{1}{(n-3)!}P_\omega(u)^{n-3}
,\quad \forall n \geq 3 \; ,
\end{equation}
where the functions $Q_\omega(m,u)$ and $P_\omega (u)$ are given by the integrals

\begin{eqnarray*}
Q_\omega(m,u) & := & \int_u^\infty (1 +|v|)^{m-4} \frac{1}{|\omega|} V_0(v)
\:dv \\
P_\omega (u) & := & 16 C_1(u_0) C_6 \int_u^\infty \frac{1}{|\omega|} V_0(v) \:dv \; ,
\end{eqnarray*}
where $C_6$ is a constant chosen such that for all $x \geq v \geq u_0$
$$ (1+|x|)^{k-m} \leq C_6 (1+|v|)^{k-m} \; , \quad 0\leq k\leq m \leq 4 \; .$$
The initial step is now given by (\ref{eq: 2.Iteration Induktionsanfang}). So assume
that (\ref{eq: Induktion der absch der ome abln}) holds for $n$. Then, according to
the iteration scheme,
\begin{eqnarray*}
\Big| \partial_\omega^m \phi_\omega^{(n+1)} (u) \Big| & \leq & \Big| \sum_{k=0}^m
\begin{pmatrix} m \\ k \end{pmatrix} \int_u^\infty C_1(u_0) (1+|v|)^{m-k}
\frac{1}{|\omega|} V_0(v) \\
& & \times 16 C_1(u_0) C_5  \frac{1}{|\omega|} Q_\omega(k,v) 
\frac{1}{(n-3)!}P_\omega(v)^{n-3} \:dv \Big| \; .
\end{eqnarray*}
Using the inequality
$$ Q_\omega (k,v) \leq C_6 (1+|v|)^{k-m} Q_\omega(m,v) $$
and the monotonicity of $ Q_\omega $, we obtain
\begin{eqnarray*}
\Big| \partial_\omega^m \phi_\omega^{(n+1)} (u) \Big| & \hspace*{-2mm}\leq &
\hspace*{-2mm} 16
C_1(u_0)
C_5 \frac{1}{|\omega|}  \int_u^\infty 16 C_1(u_0)
(1+|v|)^{m-k}\frac{1}{|\omega|} V_0(v) \\ & & \hspace*{1.5cm} \times C_6
(1+|v|)^{k-m}
Q_\omega(m,v) \frac{1}{(n-3)!}P_\omega(v)^{n-3} \:dv \\
&\hspace*{-2mm} \leq & \hspace*{-2mm} 16 C_1(u_0)
C_5 \frac{1}{|\omega|} Q_\omega(m,u) \int_u^\infty \frac{dP_\omega}{dv} (v)
\frac{1}{(n-3)!}P_\omega(v)^{n-3} \:dv \\
&\hspace*{-2mm} = & \hspace*{-2mm}16 C_1(u_0) C_5
\frac{1}{|\omega|} Q_\omega(m,u) \frac{1}{(n-2)!}P_\omega(u)^{n-2} \; ,
\end{eqnarray*} 
and (\ref{eq: Induktion der absch der ome abln}) follows. In particular, we get for
all $u\geq u_0$ and $m \leq 4$ the estimate
\begin{eqnarray} 
\Big| \partial_\omega^m \grave{\phi}_\omega (u) - \partial_\omega^m e^{-i \omega
 u} \Big| & \leq & C_5 \frac{1}{|\omega|} (1+|u|)^{m-2} + \frac{1}{|\omega|} C_5 (1 +
|u|)^{m-4} \nonumber\\ 
& & + 16 C_1(u_0) C_5
\frac{1}{|\omega|} Q_\omega(m,u) e^{P_\omega (u)} \; , \label{eq: absch der ome abln
von grave phi}
\end{eqnarray}
and the right hand side obviously tends to zero as $|\omega| \rightarrow \infty$.

In an analog way using the iteration scheme (\ref{Iterationsschema fuer phi 1}) for
$\acute{\phi}$, one shows for all $ u\leq u_0$ and $m \in\{0,...,4\}$
\begin{equation} \label{eq: absch der ome abln von acute phi}
\Big| \partial_\omega^m \acute{\phi}_\omega (u) - \partial_\omega^m e^{i\omega u
} \Big| \leq \sum_{n=1}^\infty \frac{1}{n!} M_\omega(m,u)^n = e^{M_\omega(m,u)}-1 \;
,
\end{equation}
where $M_\omega(m,u)$ is given by
$$ M_\omega(m,u) := \frac{C_7}{|\omega|} \int_{- \infty}^u (1+|v|)^m V_0(v)
\:  dv \; ,$$
with a sufficiently large constant $C_7$. Note that this integral is well defined,
and in particular the estimate is obtained easier, due to the fact that $V_0(v)$
decays exponentially as $v \rightarrow - \infty$. Moreover, the right hand side in
(\ref{eq: absch der ome abln von acute phi}) also goes to zero as $|\omega|
\rightarrow \infty$. Thus, due to (\ref{eq: absch der ome abln von grave phi}) and
(\ref{eq: absch der ome abln von acute phi}), the $\omega$-derivatives of the
fundamental solutions up to the fourth order
$\partial_\omega^m \acute{\phi}_\omega(u)$,$\partial_\omega^m\grave{\phi}_\omega(v)$
are controlled for large $|\omega|$ by constants, which depend on $u$ and the
support of the initial data $\Psi_0$. One also shows with these results and applying
the same arguments to $\acute{\phi}_\omega ',\grave{\phi}_\omega'$ that the
Wronskian $w(\acute{\phi},\grave{\phi})$ behaves as $\mathcal{O}(|\omega|)$ and
$\partial_\omega^m w(\acute{\phi},\grave{\phi}), m \leq 4$ is bounded by constants as
$|\omega| \rightarrow \infty$. Hence, interchanging in the representation (\ref{eq:
integrand in der integraldarstellung}) the differentiation with respect to
$\omega$ and the integral, which is no problem because of the compact support
of $\Psi_0$, making the substitutions
\begin{eqnarray*}
\grave{\phi}_\omega (v)\hspace*{-2mm}& =& \hspace*{-2mm}\frac{1}{\omega^2} \left( -
\grave{\phi}''_\omega(v) +
V_0(v) \grave{\phi}_\omega(v) \right)  \; ,\\
\partial_\omega \grave{\phi}_\omega (v)\hspace*{-2mm}& =& \hspace*{-2mm}
\frac{-2}{\omega^3} \left( -
\grave{\phi}''_\omega(v) + V_0(v) \grave{\phi}_\omega(v) \right) + \frac{1}{\omega^2}
\left( - \partial_\omega\grave{\phi}''_\omega(v) +
V_0(v) \partial_\omega \grave{\phi}_\omega(v) \right)
\end{eqnarray*}
as well as the analog substitutions for the second, third and
fourth $\omega$-derivative [Note that in the region $|\omega|\geq 1$
$\grave{\phi}_\omega(v)$ is $C^4$ with respect to $\omega$, cf. Lemma \ref{lemma:
bessere Entwicklung von gravephi for l=0}, $n=4$] and integrating by parts with
respect to $v$, one immediately has decay at least of $1/\omega^2$. Thus iterating
this procedure, which can be done because $V_0$
and $\Psi_0$ are smooth, yields arbitrary decay in $\omega$ and the lemma is proven.
\end{proof}

\begin{remark}
Since the method of the proof does not depend on the highest order
$\omega$-derivative, the statement of Lemma \ref{lemma: abfall der ome abln des
integranden} can be extended to arbitrary $m$. The only point where one has to be
careful is the derivation of (\ref{eq: Schranke omega abl 2.Iteration alle u}),
since for $\omega$-derivatives of higher order one has to calculate and subtract
more exact terms than in (\ref{eq: Fall l=4 absch iteration - exakter term}), due to
convergence problems. If (\ref{eq: Schranke omega abl 2.Iteration alle u}) is not
sufficient, in order to start the induction, one has to iterate this procedure
appropriately many times.
\end{remark}

We are now ready to state and prove our main theorem:

\begin{theorem} \label{theorem: Haupttheorem abfall l=0}
Consider the Cauchy problem of the scalar wave equation in the Schwarzschild
geometry
$$ \square \phi = 0\; , \quad (\phi_0, i \partial_t \phi_0)(0,r,x) = \Phi_0(r,x)$$
for smooth spherical symmetric initial data $\Phi_0 \in C^{\infty}_0 ( (2M,\infty)
\times S^2)^2 $ which is compactly supported outside the event horizon. 
Let $\Phi (t) = (\phi (t), i \partial_t \phi (t)) \in C^\infty(\R \times (2M, \infty)
\times S^2)^2$ be the unique global solution which is compactly supported for all
times $t$. Then for fixed $r$ there is a constant $c= c(r,\Phi_0)$
such that for large $t$
\begin{equation} \label{eq: decay normal}
|\phi(t)| \leq \frac{c}{t^3} \; .
\end{equation}
Moreover, if we have initially momentarily static initial data, i.e. $\partial_t
\phi_0 \equiv 0$, the solution $\phi(t)$ satisfies
\begin{equation} \label{eq: decay initially momentarily static}
|\phi(t)| \leq \frac{c}{t^4} \; .
\end{equation}
\end{theorem}

\begin{proof}
First, we decompose our initial data $\Phi_0$ into spherical harmonics. Due to the
spherical symmetry we obtain $\Phi_0(r,\vartheta, \varphi) = \tilde{\Phi}_0(r)  
Y_{00}(\vartheta,\varphi)$, where $\tilde{\Phi}_0(r) \in C^\infty_0((2M,\infty))^2$.
Introducing the Regge-Wheeler coordinate $u(r)$ and making the substitution
$\Psi(t,u) = r(u) \tilde{\Phi} (t,r(u))$, our solution
has the representation
\begin{equation} \nonumber
\Phi(t,r,\vartheta,\varphi) = \frac{1}{r} \Psi(t,u(r)) Y_{00}(\vartheta,\varphi) \; ,
\end{equation}
where $\Psi(t,u)$ satisfies
\begin{eqnarray} \nonumber
\Psi(t,u) = \hspace*{95mm}\\ -\frac{1}{\pi}\int_{\mathbb{R}} e^{-i \omega t} \left(
\int_{\mathrm{supp}\: \Psi_0} \mathrm{Im} \! \left(
\frac{\acute{\phi}_{\omega}(u)
\grave{\phi}_{\omega}(v)}{w(\acute{\phi}_{\omega },\grave{\phi}_{\omega
})}\right)
 \left(%
\begin{array}{lc}
  \omega & 1 \\
  \omega^2 & \omega \\
\end{array}%
\right)  \Psi_0(v) dv \right) d\omega,   \label{eq: lsg fall sphae symm}
\end{eqnarray}
with initial data $\Psi_0(u) := r(u) \tilde{\Phi}_0(u)$ and the Jost solutions
$\acute{\phi},\grave{\phi}$ in the case $l=0$.
According to the detailed analysis of (\ref{eq: essentieller
teil des integralkerns ohne im}) with respect to $\omega$, the term
\begin{eqnarray*}
\mathrm{Im} \! \left(
\frac{\acute{\phi}_{\omega}(u)
\grave{\phi}_{\omega}(v)}{w(\acute{\phi}_{\omega },\grave{\phi}_{\omega
})}\right) - c_0(u) g_{20}(v) \omega^2 \sign(\omega) - c_{32}(u) g_{32}(v) \omega^3
\log^2|\omega| \\
-c_{31}(u) g_{31}(v) \omega^3 \log|\omega| -c_{30}(u) g_{30}(v) \omega^3
\sign(\omega) 
\end{eqnarray*}
is $C^3(\R)$ with respect to $\omega$ for fixed $u\in \R$, $v \in \mathrm{supp
\Psi_0}$,
where the $c_{ij} (u)$, $g_{ij}(v)$ denote the appropriate coefficient functions.
[Note
that these are linearly dependent due to the symmetry of (\ref{eq: essentieller part
vom Integralkern}) with respect to $u,v$.]
Thus, defining
$$  f(\omega,u) := \left(\int_{\mathrm{supp} \: \Psi_0}  \mathrm{Im} \! \left(
\frac{\acute{\phi}_{\omega}(u)
\grave{\phi}_{\omega}(v)}{w(\acute{\phi}_{\omega },\grave{\phi}_{\omega
})}\right)
 \left(%
\begin{array}{lc}
  \omega & 1 \\
  \omega^2 & \omega \\
\end{array}%
\right)  \Psi_0(v) dv \right)_1 \; ,$$
where the subscript denotes the first vector component, the term
\begin{eqnarray*}
\tilde{f}(\omega,u) := f(\omega,u) - \Big( c_0(u) d_{20}( \psi_0^2)\: \omega^2
\sign(\omega) + c_{32}(u) d_{32}(\psi_0^2) \: \omega^3
\log^2|\omega|   \\ 
+c_{31}(u) d_{31}(\psi_0^2) \: \omega^3 \log|\omega| +c_{30}(u) d_{30}(\psi_0^2)
\: \omega^3 \sign(\omega) \Big)\eta (\omega) \\
=: f(\omega,u) -r(\omega,u) \; ,
\end{eqnarray*}
is also $C^3(\R)$ with respect to $\omega$. Here, $\psi_0^2$ denotes the second
component of the initial data $\Psi_0$,
$$d_{ij}(\psi_0^2):= \int_{\mathrm{supp} \: \Psi_0} g_{ij}(v) \psi_0^2(v) \: dv \; , 
$$ and $\eta (\omega) \in C^\infty_0(\R)$ is a smooth cutoff-function which is
identically to $1$ on a neighborhood of $\omega=0$ and $0$ outside a compact set.
Moreover, because of Lemma \ref{lemma: abfall der ome abln des integranden} the
$\partial_\omega^m \tilde{f}(\omega,u), m\in \{0,1,2,3\}$ have rapid decay for large
$|\omega|$ and are in particular $L^1(\R)$ with respect to $\omega$.
Thus, due to (\ref{eq: lsg fall sphae symm}), the first component of $\Psi$ satisfies
\begin{eqnarray*}
\psi^1(t,u) & = &  -\frac{1}{\pi} \int_\R e^{-i\omega
t}\tilde{f}(\omega,u)\:d\omega -\frac{1}{\pi} \int_\R e^{-i\omega t} r(\omega,u) \:
d\omega \\
& = &  -\frac{1}{(it)^3\pi} \left( \int_\R \tilde{f}(\omega,u)
\partial_\omega^3 e^{-i\omega
t}\:d\omega + \int_\R  r(\omega,u) \partial_\omega^3 e^{-i\omega
t} \:
d\omega \right) \; .
\end{eqnarray*} 
We write the second integral as $\int_{- \infty}^0  + \int_0^\infty$, integrate
every integral three times by parts and obtain
\begin{eqnarray*}
\psi^1(t,u) &=& \frac{1}{ (it)^3 \pi}\left(4c_0(u) d_{20}(\psi_0^2) + \int_\R e^{-i
\omega t} \partial_\omega^3 \tilde{f}(\omega,u)\:d\omega  \right. \\ 
& & \left. +\int_{-\infty}^0 e^{-i \omega t}\partial_\omega^3 r(\omega,u) \: d\omega
+ \int_0^\infty e^{-i \omega t}\partial_\omega^3 r(\omega,u) \: d\omega \right) \; .
\end{eqnarray*}
Note that the other boundary terms vanish, because the $\partial_\omega^m
\tilde{f}(\omega),m\leq3$ have rapid decay and $ \eta (\omega)\equiv
0$ outside of a compact set. Obviously, all integrals are well defined, and the
Riemann-Lebesgue lemma shows the claim in the first
case. If the initial data is initially momentarily static, all the $d_{ij}
(\psi_0^2)$ vanish and the entries in the matrix in (\ref{eq: lsg fall sphae
symm}) yield an additional $\omega$. Hence, the highest irregular term is
$c_0(u) d_{20}(\psi_0^1) \omega^3 \sign(\omega)$, and the same arguments as before
conclude the proof.
\end{proof}

\begin{remark}
The decay rates $1/t^3$, and $1/t^4$, respectively, are optimal in the sense that
there exists initial data such that these cannot be improved. This is obvious due to
the fact that $c_0(u)>0$.
\end{remark}

\section{Discussion on the case $l \neq0$}

According to Price's Law \cite{Price}, the $lm$-component $\Phi^{lm}(t,u) =
\frac{1}{r} \Psi^{lm}(t,u)$ of a solution for the Cauchy
problem for the scalar wave equation in Schwarzschild spacetime with
compactly supported smooth
initital data generally falls off at late times $t$ as
$t^{-2l -3}$ and $t^{-2l-4}$ for initially momentarily static initial data,
respectively. This has been confirmed in the previous section for spherical
symmetric initial data, i.e. in the case $l=0$ [cf. Theorem \ref{theorem:
Haupttheorem
abfall l=0}]. Moreover, there is numerical evidence which lets us conjecture this to
be correct \cite{Kar}.
We briefly discuss whether the methods of the preceding section
still apply to the case when the angular mode $l$ is non-zero.

To this end, let us reconsider the
construction of the fundamental solutions $\grave{\phi}_{\omega l}$ of the
Schr\"odinger equation (\ref{Schroedinger equation}).
First, we
make some remarks about the fundamental solutions $ \omega^l \grave{\phi}_{\omega l}
(u)$ (see also \cite[Section 5]{Kronthaler}). The fundamental solutions were
constructed as the series 
\begin{eqnarray} \label{eq: Reihenansatz grave phi l neq 0}
\omega^l \grave{\phi}_\omega (u) = \sum_{m=0}^\infty \phi_\omega^{(m)} (u) \; ,
\end{eqnarray}
where the $\phi^{(m)}$ are given by the iteration scheme
\begin{eqnarray} \label{eq: Iterationsschema grave phi l neq 0}
\phi_\omega^{(m+1)} (u) = - \int_u^\infty S_\omega(u,v) W_l(v) \phi_\omega^{(m)}(v)
\:dv \; ,
\end{eqnarray}
with potential, cf. also Lemma \ref{lemma: asymptotische Entwicklung von V_l},
\begin{eqnarray} \label{eq: restpotenzial l neq 0}
W_l(u) = V_l(u) - \frac{l(l+1)}{u^2} & = & c_{31} \frac{\log u}{u^3} + 
\frac{c_{30}}{u^3} + h(u) \; , \\ \textrm{where \ } h(u) & = & \mathcal{O} \left(
\frac{\log^2
u}{u^4} \right) \nonumber
\end{eqnarray}
for large $u$, and Green's function
\begin{eqnarray} \label{eq: Greenskern l neq 0 teil 1}
  S_\omega(u,v) = \frac{(-1)^{l+1}}{\omega} \big(h_1(l,\omega v)
h_2(l,\omega u) - h_1(l, \omega u) h_2(l, \omega v)\big) \; ,
\end{eqnarray}
where 
\begin{eqnarray} \label{eq: Greenskern l neq 0 teil 2}
h_1(l, \omega u) = \sqrt{\frac{\pi \omega u}{2}} J_{l+1/2} (\omega u) \; , \quad
h_2(l, \omega u) =  \sqrt{\frac{\pi \omega u}{2}} J_{-l-1/2} (\omega u) \; ,
\end{eqnarray}
and $J_\nu$ denotes the Bessel function of the first kind.
As initial function $\phi_\omega^{(0)}(u)$ we have chosen 
\begin{equation*} 
\phi_\omega^{(0)} (u) = \omega^l e^{-i (l+1) \frac{\pi}{2}} \sqrt{\frac{\pi
\omega u}{2}} H_{l+ 1/2}^{(2)}(\omega u) \; ,
\end{equation*}
where $H^{(2)}_\nu$ denotes the second Hankel function. Since $l$ is an
integer, these functions are directly connected to the spherical Bessel functions and
simplify significantly. Namely, $h_1,h_2$ have the following representations [cf.
\cite[Chapter 10]{AS}]
\begin{eqnarray}
\label{eq: Polynomialdarstellung fuer spherical Bessel functions l+1/2}
h_1 (l, \omega u)  =  P(l+\frac{1}{2}, \omega u) \sin(\omega u- \frac{1}{2} l \pi)
+ Q (l+\frac{1}{2}, \omega u) \cos (\omega u - \frac{1}{2} l \pi) \; \;\\
\label{eq: Polynomialdarstellung fuer spherical Bessel functions -l-1/2}
h_2 (l, \omega u)  =  P(l+\frac{1}{2}, \omega u) \cos(\omega u+ \frac{1}{2} l \pi)
- Q (l+\frac{1}{2}, \omega u) \sin (\omega u + \frac{1}{2} l \pi) \; \;
\end{eqnarray}
where $P,Q$ are finite polynomials given by
\begin{eqnarray*}
P(l+ \frac{1}{2}, \omega u)&=& \sum_{k= 0}^{[\frac{1}{2} l]} \: (-1)^k 
\frac{(l+\frac{1}{2},2k)}{(2 \omega u)^{2k}} \; ,\\
Q(l+ \frac{1}{2}, \omega u)&=& \sum_{k=0}^{[\frac{1}{2} (l-1)]} (-1)^k 
\frac{(l+\frac{1}{2},2k+1)}{(2 \omega u)^{2k+1}} \; ,
\end{eqnarray*}
with $$ (l+ \frac{1}{2},k) = \frac{(l+k)!}{k! \: \Gamma(l-k+1)} \; .$$
And the initial function can be expressed by
\begin{equation}
\label{eq: Polynomialdarstellung fuer anfangsfunktion}
\phi_\omega^{(0)} (u) = \omega^l e^{-i \omega u} \sum_{k=0}^l
\frac{(l+\frac{1}{2},k)}{(2 i \omega u)^k} \; .
\end{equation}
Due to the recurrence formulas for the derivatives of the Bessel functions, we have
the identities
\begin{eqnarray*}
\partial_\omega h_1(l, \omega u)  =  u h_1(l-1, \omega u) - \frac{l}{\omega}
h_1(l,\omega u) \hspace*{2mm} \; , \\ \partial_\omega h_2(l, \omega u)  =  - u
h_2(l-1, \omega u)
- \frac{l}{\omega} h_2(l,\omega u) \; .
\end{eqnarray*}
As a consequence,
\begin{eqnarray*}
\partial_\omega S_\omega (u,v) = & \hspace*{-2mm} - \hspace*{-2mm} &
\frac{2l+1}{\omega} S_\omega
\\
 &\hspace*{-2mm}+\hspace*{-2mm}&v \frac{(-1)^{l+1}}{\omega} (h_1(l-1,\omega v)
h_2(l,\omega u) + h_1(l,\omega
u) h_2(l-1, \omega v) )  \\
 &\hspace*{-2mm}+\hspace*{-2mm}&u \frac{(-1)^l}{\omega} (h_1(l,\omega v)
h_2(l-1,\omega u) + h_1(l-1,\omega
u) h_2(l, \omega v) ) \; . \hspace*{1.3mm}
\end{eqnarray*}
This allows us to derive the necessary estimates for the Green's function $S_\omega
(u,v)$.
Exploiting the asymptotics we have already
seen in \cite[Section 5]{Kronthaler}
\begin{equation*} 
| S_\omega (u,v) | \leq C_1 \left(\frac{u}{1 + |\omega| u} \right)^{-l}
\left(\frac{v}{1 + |\omega| v} \right)^{l + 1} e^{v| \Im \omega|
 +u \Im \omega} \; ,
\end{equation*}
for $v\geq u >0$ and an appropriate constant $C_1$.
In order to derive an estimate for $\partial_\omega S_\omega$ and small $|\omega|$,
we make use of
\begin{eqnarray*}
h_1(l,\omega u) & \sim & k_1 (\omega u)^{l+1} + k_2 (\omega u)^{l+3}  \\
h_2(l,\omega u) & \sim & k_3 (\omega u)^{-l} + k_4 (\omega u)^{-l+2} \; ,
\quad  \textrm{if} \; |\omega| u \ll 1 \; ,
\end{eqnarray*}
and certain constants $k_1,...,k_4$ [refer to the series expansion of the Bessel
functions \cite[9.1.10]{AS}] to obtain (note that $v\geq u >0$),
\begin{eqnarray*}
| \partial_\omega S_\omega (u,v)| \leq C_2 \left(\frac{u}{1 + |\omega|
u} \right)^{-l}
\left(\frac{v}{1 + |\omega| v} \right)^{l + 2} \; , & \textrm{if } |\omega|
v \ll 1 \; .
\end{eqnarray*}
For large arguments $|\omega| u \gg 1$ we use (\ref{eq:
Polynomialdarstellung fuer spherical Bessel functions l+1/2}),(\ref{eq:
Polynomialdarstellung fuer spherical Bessel functions -l-1/2}) and get by a
straightforward calculation
\begin{eqnarray*}
\partial_\omega S_\omega (u,v) \sim \frac{-2l}{\omega^2} \sin(\omega(u-v)) +
\partial_\omega \left[ \frac{1}{\omega} \sin(\omega(u-v)) \right] \; , & \textrm{if }
| \omega | u \gg 1 .
\end{eqnarray*}
Together with (\ref{eq: abschaetzung der omegaabl der greensfkt}), we obtain
\begin{eqnarray*} 
| S_\omega (u,v) | \leq C_3 \frac{v^2}{1 + |\omega| v} e^{v| \Im
\omega| +u \Im \omega} \; , & \textrm{if } | \omega | u \gg 1.
\end{eqnarray*}
Combining these estimates, we find a constant $C$ such that
\begin{equation} \label{eq: Abschaetzung fuer partial omega S_omega}
| \partial_\omega S_\omega (u,v) | \leq C \left(\frac{u}{1 + |\omega| u}
\right)^{-l}
\left(\frac{v}{1 + |\omega| v} \right)^{l + 1} v \: e^{v| \Im \omega|
 +u \Im \omega} \; ,
\end{equation}
for $v\geq u >0$.
Moreover, looking at (\ref{eq: Polynomialdarstellung fuer anfangsfunktion}) we get
the following bounds for the initial function, 
\begin{eqnarray}
\label{eq: Abschaetzung fuer anfangsfkt}
| \phi_\omega^{(0)} (u) | & \leq & C_4  \left(
\frac{u}{1+ | \omega| u} \right)^{-l} e^{u \Im \omega } \; , \\
\label{eq: Abschaetzung fuer partial omega anffkt} 
| \partial_\omega \phi_\omega^{(0)} (u) | & \leq & C_5  \left(
\frac{u}{1+ | \omega| u} \right)^{-l} u \: e^{u \Im \omega } \; .
\end{eqnarray}
These estimates allow us to proceed in exactly the same way as in the proof of Lemma
\ref{lemma: Entwicklung von acute phi,l=0}. As analogon to
$\hat{\phi}^{(1)}_\omega (u)$ we obtain the term
$$ - \int_u^\infty S_\omega(u,v) \left( c_{31} \frac{\log v}{v^3} +
\frac{c_{30}}{v^3} \right) \phi_\omega^{(0)} (v) \:dv,$$
which we calculate using (\ref{eq: Polynomialdarstellung fuer spherical Bessel
functions l+1/2}),(\ref{eq: Polynomialdarstellung fuer spherical Bessel functions
-l-1/2}) and (\ref{eq: Polynomialdarstellung fuer anfangsfunktion}). Essentially, we
get integrals of the shape
$$ \frac{\omega^l}{(\omega u)^n \omega^{m+k+1}} \left(C_6 e^{i \omega u}
\int_u^\infty e^{-2 i \omega v} \frac{\log^q v}{v^{3+k+m}}\:dv + C_7e^{-i \omega u}
\int_u^\infty \frac{\log^q v}{v^{3+k+m}} \: dv \right)\; ,$$
where $q \in \{0,1\}, 0 \leq n,m,k \leq l$. Note that the terms involving $\omega$
singularities resolve, due to the fact that $\omega^l \grave{\phi}_\omega$ is
continuous with respect to $\omega$. Computing these integrals via Lemma \ref{lemma:
zur Reihenentwicklung} (in the limit $\varepsilon \rightarrow 0$), we see (as before)
that the only terms not being $C^1$ with respect to $\omega$ are of the form
\begin{equation} \label{eq: erste Irregularitaet im Fall l neq 0}
e^{i\omega u} \frac{1}{\omega^{m+k+1}} (2i \omega)^{k
+m+2}  \left( \log^2(2 i \omega u) + \log u \log(2i \omega u) +\log(2i \omega
u)\right) \; ,
\end{equation} modulo coefficients.
Now, we apply the same iteration with analog estimates and all in all we have shown:
\begin{lemma} \label{lemma: Entwicklung von acute phi,l geq 1}
For $l\geq 1$, $\omega \in \R \setminus \{0\}$ and fixed $u>0$ the fundamental
solutions
$\omega^l \grave{\phi}_\omega(u)$ have the representation
\begin{eqnarray} 
\nonumber \omega^l \grave{\phi}_\omega(u)= \phi_\omega^{(0)}(u) + g_3(\omega,u) + 2 i
\omega \log^2(2 i \omega) g_4(\omega,u) \hspace*{4.7mm} \\ +2 i \omega \log(2 i
\omega) g_5(\omega,u) +2 i \omega g_6 (\omega,u)\; ,  \label{eq: Entwicklung von
acute phi,l geq 1}
\end{eqnarray} 
where the functions $g_3,g_4,g_5$ and $g_6$ are $C^1(\R)$ with respect to $\omega$.
\end{lemma} 

Hence,
we still have finite expressions for the Green's function $S_\omega(u,v)$ as well as
for
the initial function $\phi_\omega^{(0)}(v)$, which involve essentially the plane
waves $e^{ \pm i \omega u},e^{\pm i \omega v}$. Expanding all these expressions and
deriving estimates analog to (\ref{eq: Abschaetzung fuer partial omega S_omega}) and
(\ref{eq: Abschaetzung fuer partial omega anffkt}) for higher order
$\omega$-derivatives, we can improve Lemma \ref{lemma: Entwicklung von acute phi,l
geq 1} in the same way as Lemma \ref{lemma: Entwicklung von acute phi,l=0}
following the arguments of the proof of Lemma \ref{lemma: bessere Entwicklung von
gravephi for l=0}.
Also, a similar result to Corollary \ref{corollary: Beziehung g_ijk g_0} seems
straightforward. The problem now arises, when we have to derive an $\omega-$expansion
of the essential part of the integral kernel 
\begin{equation} \label{eq: essentieller teil des Integralkerns l neq 0}
\mathrm{Im} \left(\frac{\acute{\phi}_{\omega l} (u) \grave{\phi}_{\omega
l}(v)}{w(\acute{\phi}_l,\grave{\phi}_l)} \right) \; .
\end{equation}
The main difficulty can be seen as follows. If we proceeded in the same way as in
the case $l=0$, the lowest regular term
with respect to $\omega$ should appear with the power $\omega^{2l+2}$ [cf.
proof of
Theorem \ref{theorem: Haupttheorem abfall l=0}] in order to satisfy Price's law. But
due to the fact that the first irregularity in $\omega$ looks as follows,
$$e^{i \omega u}u^{-l} 2i\omega \big(c\log^2 (2i\omega u)+c \log u \log(2i\omega u) +
c\log(2i\omega u)\big) \; ,$$ 
[cf. equation (\ref{eq: erste Irregularitaet im Fall l neq 0})],
we would have to find a systematic way in order to check that the coefficients in
front of terms with lower
regularity vanish. Because of the complexity of the calculations we did not succeed
in this point. Thus, following
the same arguments as for $l=0$ together with the analog result to Corollary
\ref{corollary: Beziehung g_ijk g_0}, which would involve $2 i \omega
\log^2(2i\omega)$ as highest irregularity, we would have to assume $ \omega
\log|\omega|$ as the lowest regular term in the expansion of (\ref{eq: essentieller
teil des Integralkerns l neq 0}). Except for this problem, we do not expect any
further difficulties in extending Lemma \ref{lemma: abfall der ome abln des
integranden} to $l \neq 0$, apart
from the complexity of the calculations and the estimates. Thus, for arbitrary $l$ it
follows a similar statement to Theorem \ref{theorem: Haupttheorem abfall l=0}, but
with the decay $|\phi(t)|\leq c/t^2$, and in the case of momentarily static initial
data $|\phi(t)| \leq c/t^3$, respectively. The proof uses essentially the arguments
of
the proof of Theorem \ref{theorem: Haupttheorem abfall l=0}, with the difference that
one basically has to check the inequality
$$ \Big|\int_{-1}^1 \log|\omega| e^{-i\omega t} \: d\omega \Big| \leq \frac{c}{t}
\: .$$
To this end, one makes the substitution $z = \omega t$ and splits up the integrals
to obtain
\begin{eqnarray*}
\int_{-1}^1 \log|\omega| e^{-i\omega t} \: d\omega & = & \frac{1}{t} \bigg(
\int_{-1}^1 \log|z| e^{-i z} \: dz - \log t \int_{-t}^t e^{-iz} \: dz  \\
& & + \int_{-t}^{-1} \log (-z) e^{-iz} \:  dz + \int_1^t \log z e^{-iz} \: dz \bigg)
\; .
\end{eqnarray*}
Computing the second integral and integrating the last two integrals by parts yields
$$ = \frac{1}{t} \bigg( \int_{-1}^1 \log|z| e^{-i z} \: dz + \frac{1}{i}
\int_{-t}^{-1} \frac{1}{z} e^{-iz} \: dz + \frac{1}{i}
\int_{1}^{t} \frac{1}{z} e^{-iz} \: dz  \bigg) \; ,$$
and the inequality follows, after having integrated the last two integrals once
again by parts followed by standard integral estimates. However, in view of Price's
law, this result is not
satisfying.

\begin{acknowledgment}
The author would like to thank Felix Finster, University of Regensburg, for
introducing to this problem and also for his helpful suggestions and remarks.
\end{acknowledgment}

\noindent
NWF I -- Mathematik,
Universit{\"a}t Regensburg, 93040 Regensburg, Germany, \\
{\tt{Johann.Kronthaler@mathematik.uni-regensburg.de}}

\end{document}